\documentclass{cmspaper}
\usepackage{graphicx}
\usepackage{floatflt}
\usepackage{color}
\usepackage{colortbl}
 
\newcommand{\Pt}{{P_t}}
\newcommand{\Et}{{E_t}}
\newcommand{\dphi}{\Delta\phi}
\newcommand{\phizj}{\phi_{(Z,jet)}}

\newcommand{\ptzj}{$~\Pt^Z$ and $\Pt^{Jet}~$}
\newcommand{\PTzj}{$~\Pt^Z-\Pt^{Jet}~$}

\newcommand{\la}{\langle}
\newcommand{\ra}{\rangle}
\newcommand{\zpj}{~``$Z^0+jet$''~}
\newcommand{\gpj}{~``$\gamma+jet$''~}
\newcommand{\rrr}{\to} 

\newcommand{\pth}{\hat{p}_{\perp}^{\;min}}
\newcommand{\Db}{\Pt(O\!+\!\eta>5)}
\newcommand{\Ptz}{\Pt^{Z}}

\newcommand{\ptz}{$\Pt^{Z}$~}
\newcommand{\ptg}{$\Pt^{\gamma}$}
\newcommand{\ptj}{$\Pt^{jet}$}
\newcommand{\Fptzj}{(\Pt^{Z}-\Pt^{jet})/\Pt^{Z}}

\newcommand{\ZJ}{(\Pt^{Z}-\Pt^{J})/\Pt^{Z}}
\newcommand{\Zpart}{(\Pt^{Z}\! \!-\! \! \Pt^{part})/\Pt^{Z}}
\newcommand{\Jpart}{(\Pt^{J}\! \!-\! \! \Pt^{part})/\Pt^{J}}

\newcommand{\hmm}{\hspace*{-1.3mm}}
\newcommand{\aaa}{\hspace*{.39cm}}

\newcommand{\lt}{\!<\!}
\newcommand{\gt}{\!>\!}

\newcommand{\coltab}{0.96}

\suppressfloats[!]

\begin{document}

\begin{titlepage}



\vskip 20mm
~\\[10mm]

 \title{On the application of ``$Z^0+jet$'' events for 
  setting the absolute jet energy scale  and determining the gluon distribution 
in a proton at the LHC.}

  \begin{Authlist}
   D.V.~Bandurin, N.B.~Skachkov
    \Instfoot{jinr}{Joint Institute for Nuclear Research, Dubna, Russia}
  \end{Authlist}


~\\[10mm]

\begin{abstract}
A possibility of jet energy scale setting by help of 
``$p p\to Z^0+jet+X$'' process at LHC  is studied.
The effect of new set of cuts, proposed in our previous works, on the improvement
of \PTzj balance is demonstrated. The distributions of the selected events over 
\ptz ~and $\eta^{jet}$ are presented. A possibility of background events
suppression by use of the \zpj events selection criteria is shown.

It is also found that the samples of \zpj events, gained with the cuts for the jet energy calibration,
may have enough statistics for determining the gluon distribution inside a proton in the region 
of $2\cdot 10^{-4}\leq x \leq 1.0$ with $0.9\cdot10^3\leq Q^2\leq 4\cdot 10^4 ~(GeV/c)^2$.

Monte Carlo events produced by the PYTHIA 5.7 generator are used here.
\end{abstract}

\end{titlepage}

\setcounter{page}{2}

\thispagestyle{empty}

\setcounter{page}{1}
\section{Introduction.} 

A precise reconstruction of the jet energy is the extremely important task
in many  high energy physics experiments. The previous studies of possibilities
to apply for this aim different physical processes (like "$Z^0/\gamma+jet$" and others),
done in D0, CDF, CMS and ATLAS collaborations may be found in \cite{1}--\cite{BKS_P5}.

\zpj events with one high-$\Pt$ jet can provide a useful sample to perform 
{\it in situ} determination of a jet transverse momentum via 
a transverse momentum of $Z^0$ boson reconstructed from the precisely measured
leptonic $Z^0$ decay ($Z^0\to \mu^+\mu^-, e^+e^-$). 

In this paper we limit our consideration to $Z^0\to \mu^+\mu^-$ decay only.
The amount of material in front and inside the muon detector system guarantees 
absorbing most hadronic background. Besides, by using the track segments matching
between the muon system and the tracker one can reach a high enough reconstruction
efficiency of a muon track with a good momentum resolution (of order of $0.5-1\%$)
\footnote{see \cite{MS}}.
A selection of \zpj events with the consequent decay $Z^0\to e^+e^-$ would require
a supplementary introduction of isolation criteria for $e^\pm$ tracks to perform
a confident reconstruction of $e^\pm$ signal in the cells of 
electromagnetic calorimeter (ECAL)
\footnote{For instance, by requiring (1) a total transverse momentum $\Et^{tot}$ around 
an electron
with $\Et^e$ in the cone with $R=0.3$ to be $\Et^{tot}\lt5~GeV$ and (2) $\Et^{tot}/\Et^e\lt0.1$ 
(e.g. see \cite{CMJ}) we additionally reduce a number of signal 
events by $2-4\%$.}.
Our study has shown that a background to the \zpj events with $e^+e^-$ decay channel of
$Z^0$ boson is about the same as one to the \zpj events with $\mu^+\mu^-$ decay channel.

\zpj events is a useful tool to cross-check a setting an absolute jet energy scale
with help of other processes like \gpj \cite{BKS_P1}--\cite{BKS_P5} and 
``$W\!\to \! 2~jets$'' events \cite{SAV}, for example.

Here we present results of the analysis of \zpj events generated by using
PYTHIA 5.7 Monte-Carlo event simulation package \cite{PYT}.

Section 2 is an introduction into the problem. General features of \zpj processes at LHC energies 
as well as the sources of the disbalance between transverse momenta of $Z^0$ and jet
are presented here. We list here a set of the selection cuts used to identify signal \zpj events
(implying subsequent $Z^0$ decay to the muon pair). New criteria (introduced for the first time
in \cite{BKS_P1}) of $\Pt$ activity suppression beyond the \zpj system 
%
%
are also described there.
Vector and scalar forms of the balance equation of the event as a whole are given in this section.

In Section 3 we briefly describe an estimation of non-detectable part of jet $\Pt$ without 
taking into account of the detector effects.

Section 4 is devoted to the study of the influence of $\Pt^{clust}_{CUT}$ and $\Pt^{out}_{CUT}$
as well as the angle between $\vec{\Pt}^Z$ and $\vec{\Pt}^{jet}$ on the initial state radiation 
(ISR) suppression. The rates of \zpj events with jet covering Barrel, Endcap and Forward parts of
the calorimeter are also given in this section.

In Sections 5 and 6 we confine ourselves by consideration of \zpj events with the jet 
entirely contained in the Barrel region. The dependences of various physical variables on 
$\Pt^{clust}_{CUT}$ and 
$\Pt^{out}_{CUT}$ are analyzed  there and shown in the tables of Appendices 2--5.
 The values of the disbalance between \ptzj with 
for three \ptz intervals and various $\Pt^{clust}_{CUT}$ and $\Pt^{out}_{CUT}$ values are presented 
in Appendix 6.

In Section 7 we study a possibility of background events suppression for different 
\ptz intervals.

The number of events for determination of the gluon density  in a proton by using
\zpj events is estimated in Section 8. The event rates and contributions of various processes 
are calculated there for different $x$ and $Q^2$ intervals. It is shown that
the kinematic region for the gluon density determination in the intervals:
$2\cdot 10^{-4}\leq x \leq 1.0$ with $0.9\cdot10^3\leq Q^2\leq 4\cdot 10^4 ~(GeV/c)^2$
can be covered by studying the \zpj events.

\newpage
\section{Generalities of the \zpj process.}
\subsection{Leading order picture and sources of \ptzj disbalance.}        

In this section we observe briefly the main effects that lead to the disbalance
between \ptzj
\footnote{The more detailed consideration is given in our papers \cite{BKS_P1,GPJ_D0}
devoted to the jet energy calibration by using \gpj events.}.

The process of $Z^0+jet$ production
\begin{equation}
pp\rightarrow Z^0\, +\, 1\,jet\, + \,X
\label{eq:zpj}
\end{equation}
is caused at the parton level  by two subprocesses: Compton-like scattering\\[-5mm]
\begin{eqnarray}
\hspace*{6.34cm} qg\to q+Z^0 \hspace*{72mm} (2a)
\nonumber
\end{eqnarray}
\vspace{-3mm}
and the annihilation process\\[-2mm]
\begin{eqnarray}
\hspace*{6.32cm} q\overline{q}\to g+Z^0.  \hspace*{71mm} (2b)
\nonumber
\end{eqnarray}
Some leading order Feynman diagrams of these processes are shown in 
Fig.~\ref{fig:LO}. \\[-42mm]
\begin{center}
\begin{figure}[h]
  \hspace{10mm} \includegraphics[width=13cm,height=67mm]{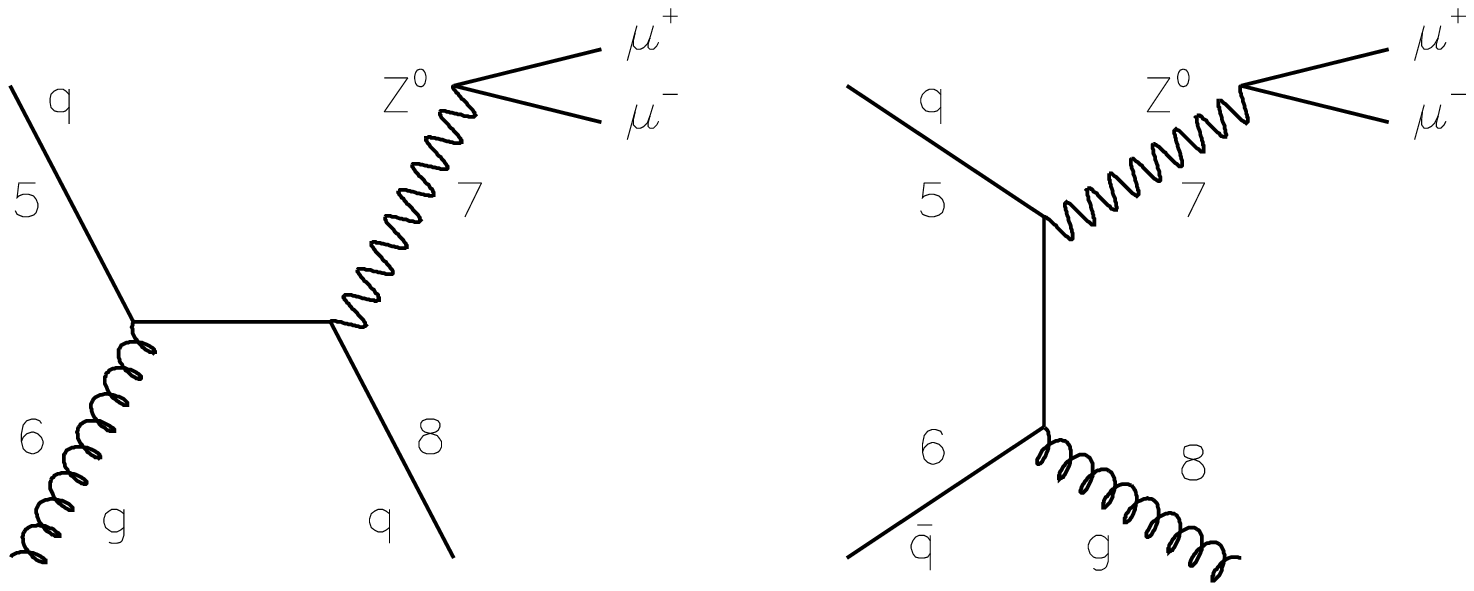} 
  \vspace{-11mm}
\caption{\hspace*{0.0cm} Some leading order Feynman diagrams for $Z^0$
production.} 
\label{fig:LO}
\end{figure}
\vskip-8mm
\end{center}

If the initial state radiation (ISR) is absent, the total transverse momentum of 
the final state in the subprocesses (2a) or (2b) is equal to zero, i.e. the $\Pt$ 
balance equation for $Z^0$ and final parton would look as\\[-12pt]
\begin{eqnarray}
\vec{\Pt}^{Z+part}=\vec{\Pt}^{Z}+\vec{\Pt}^{part} = 0.
\end{eqnarray}
Thus, having neglected hadronization effect we could expect that a jet transverse momentum
\ptj ~is close enough to $Z^0$ boson transverse momentum, i.e. 
$\vec{\Pt}^{jet}\approx-\vec{\Pt}^{Z}$.

\begin{center} 
\begin{figure}[h]
  \hspace{15mm} \includegraphics[width=13cm,height=67mm,angle=0]{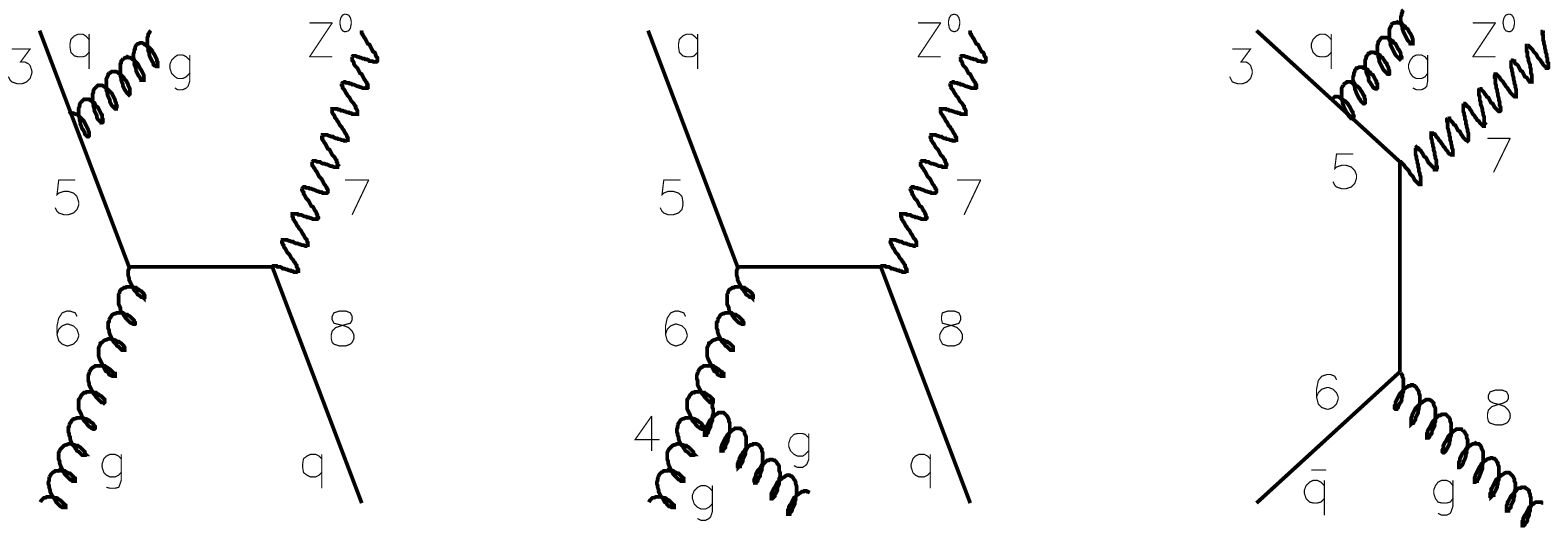}
  \vspace{-10mm}
  \caption{\hspace*{0.0cm} Some Feynman diagrams of $Z^0$ 
production including gluon radiation in the initial state.}
    \label{fig:NLO}
  \end{figure}
\end{center}

A radiation of a gluon in the initial state with a non-zero transverse momentum
$\Pt^{gluon}\equiv \Pt^{ISR}\neq 0$ can produce a disbalance between $\Ptz$ and
$\Pt^{part}$ and, thus, between  transverse momenta of $Z^0$ boson and the jet 
originated from this proton. 
The corresponding next-to-leading order diagrams are shown in Fig.~\ref{fig:NLO}.

Following \cite{BKS_P1}, we choose the sum of the modulus of
the transverse momentum vectors $\vec{\Pt}^{5}$ and $\vec{\Pt}^{6}$
of the incoming  (into $2\rrr 2$ fundamental QCD subprocesses $5+6\to 7+8$)
partons (lines 5 and 6 in Fig.~2): 
\\[-15pt]
\vspace{-1mm}
\begin{eqnarray}
\qquad \Pt{56}=|\Pt^5|+|\Pt^6|
\end{eqnarray}
as a quantitative measure to estimate the $\Pt$ disbalance caused by ISR. 

The numerical notations in the Feynman diagrams shown in Figs.~1 and 2
and in formula (2)  are chosen to be in correspondence with those
used in the PYTHIA event listing for description of the parton--parton subprocess
displayed schematically in Fig.~3. The ``ISR'' block describes the initial
state radiation process that can take place before the fundamental
hard $2\to 2$ process.
\\[-43mm]
\begin{center}
  \begin{figure}[h]
  \vspace{-12mm}
   \hspace{3cm} \includegraphics[width=8cm,height=40mm,angle=0]{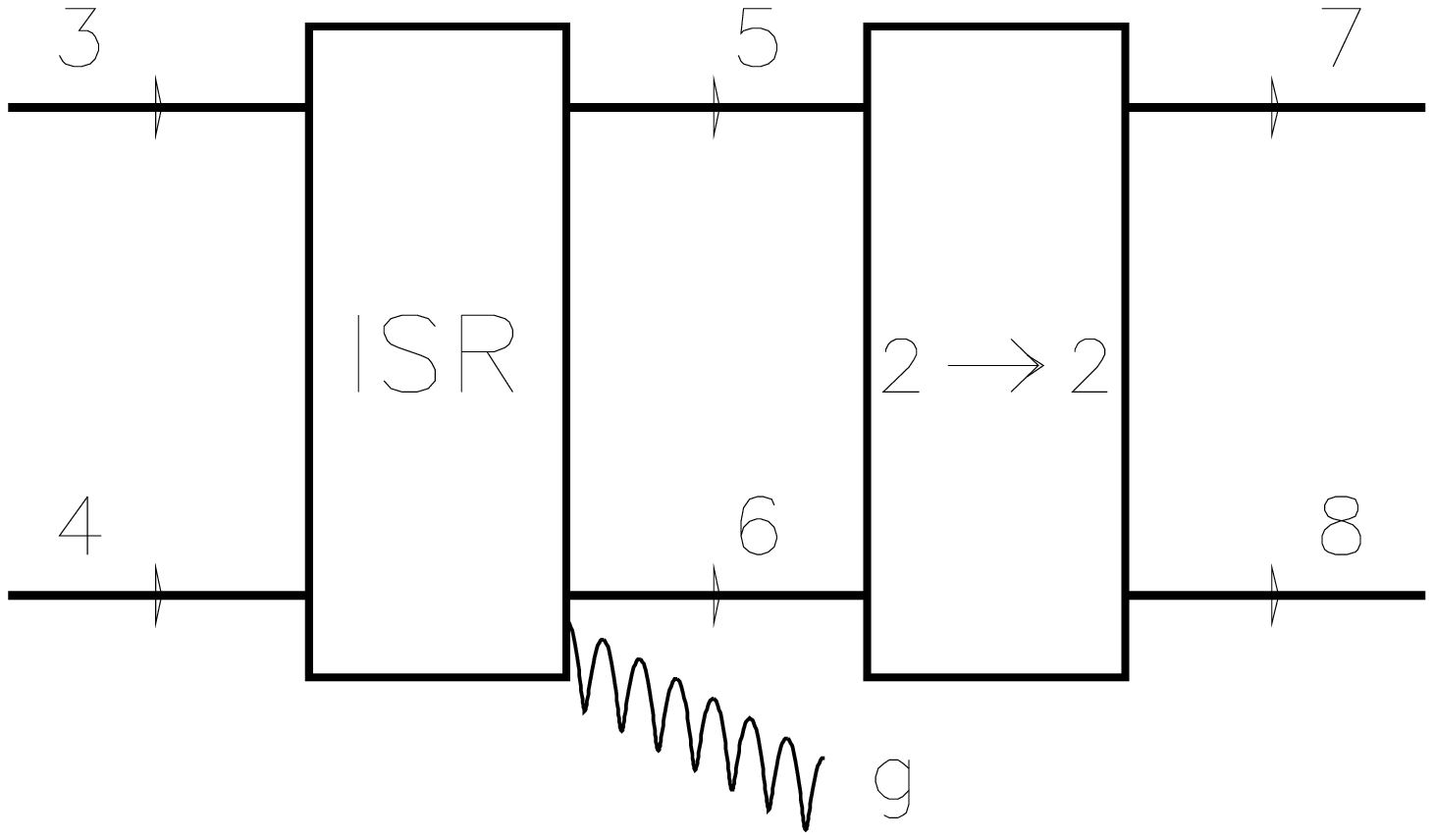}
  \vspace{-22mm}
 \caption{\hspace*{0.0cm} PYTHIA ``diagram'' of~ a fundamental $2\to2$ process (5+6$\to$7+8)
following the block (3+4$\to$5+6) of initial state radiation (ISR).}
    \label{fig4:PYT}
  \vspace{-0mm}
  \end{figure}
\end{center}

\begin{center}
\begin{figure}[htbp]
  \vspace{-42mm}
  \hspace{15mm} \includegraphics[width=13cm,height=67mm,angle=0]{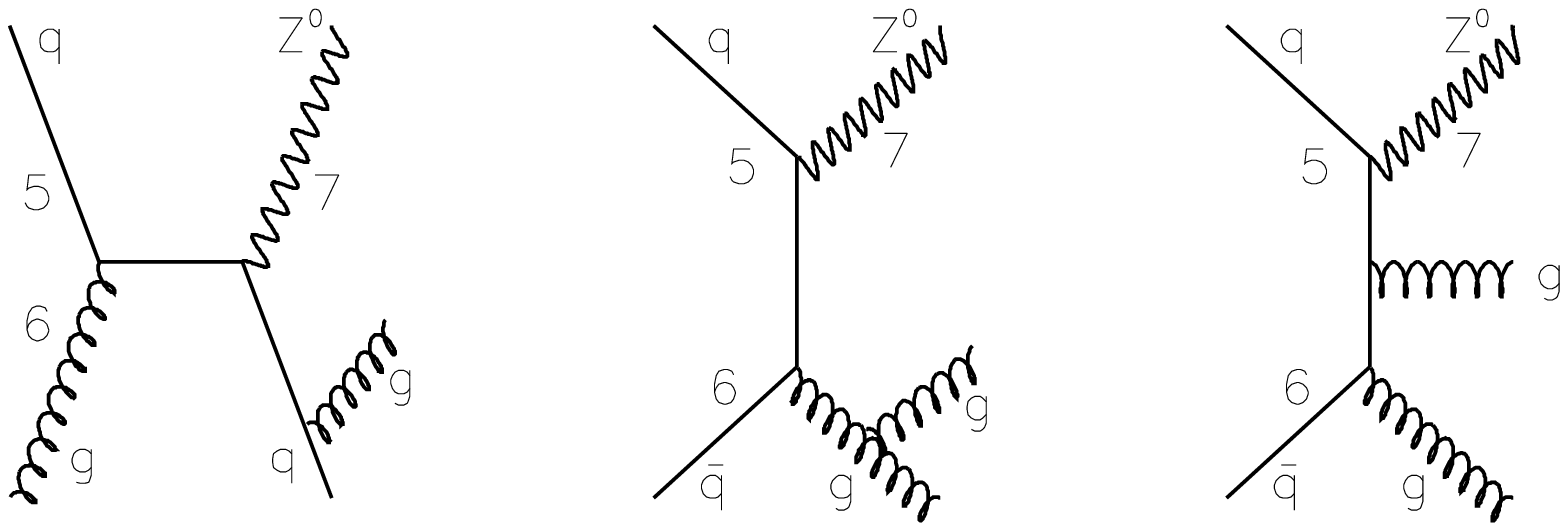}
  \vspace{-12mm}
  \caption{\hspace*{0.0cm} Some of Feynman diagrams of $Z^0$ 
production including gluon radiation in the final state.}
    \label{fig4:NLO}
  \vspace{2mm}
  \end{figure}
\end{center}

Let us consider fundamental subprocesses in which there is
no initial state radiation but instead  final state radiation
(FSR) takes place.
Some Feynman diagrams of the signal subprocesses with the FSR
are shown in Fig.~\ref{fig4:NLO}. An appearance of a gluon in
the final state may also cause a disbalance between transverse momenta
of $Z^0$ and jet. But because it manifests itself as some extra jets or
clusters, like in the case of ISR, the same selection criteria (see below) 
as for suppression of ISR can be used.


\vskip-9mm
A possible non-zero value of the intrinsic transverse momentum of a parton
inside a colliding proton ($k_T$) may be another source of the
$\Ptz$ and $\Pt^{part}$ disbalance in the final state.
Its reasonable value is supposed to lead to the value of $k_T\leq1.0 ~GeV/c$. 
In what follows we shall keep the value of $k_T$ to be fixed  by the
PYTHIA default value $\langle k_T \rangle=0.44~ GeV/c$. 
The dependence of the disbalance between $\Ptz$ and $\Pt^{Jet}$
on a possible variation of $k_T$ is discussed in detail in \cite{BKS_P5,GPJ_D0}.
 The general conclusion is that the variation of $k_T$ within
 reasonable boundaries does not produce a large effect when the initial state radiation is
taken into account. The latter makes a dominant contribution.


Another non-perturbative effect that results in the \ptzj disbalance
is an hadronization of the parton, produced in the fundamental
$2\to 2$ subprocess, into a jet. The contribution of the hadronization to this
disbalance is calculated within the Lund string fragmentation scheme
used by default in PYTHIA. The mean values of the relative $\Pt^{Jet}-\Pt^{part}$ disbalance
are presented in the tables of Appendices 2 -- 5 for three different
jetfinders (UA1, UA2 and LUCELL
\footnote{UA1 and UA2 algorithms are taken from the CMSJET program of fast simulation \cite{CMJ}.
while LUCELL is the PYTHIA's default jetfinding algorithm \cite{PYT}.}
) 
as a function of the variable which limit a cluster
activity beyond the \zpj system (see Section 2.2 and \cite{BKS_P3}).

\subsection{Definition of selection cuts.}                                                                  

\noindent
1. We shall select the events with $Z^0$ boson 
\footnote{Here and below in the paper speaking about $Z^0$ boson we imply a signal
reconstructed from the muon pair with muons selected by the criteria $2-4$ of this section.}
and  one jet with\\[-5pt]
\begin{equation}
\Pt^{Z} \geq 40~ GeV/c~ \quad {\rm and} \quad \Pt^{Jet}\geq 30 \;GeV/c.
\label{eq:sc1}
\end{equation}

For most of our applications the jet is defined according to the PYTHIA
jetfinding algorithm LUCELL.
The jet cone radius R in the $\eta-\phi$ space counted from the jet initiator cell (ic) is
taken to be $R_{ic}=((\Delta\eta)^2 + (\Delta\phi)^2)^{1/2}=0.7$.
Comparison with the UA1 and UA2 jetfinding algorithms 
is presented in Sections 5 and 6.

\noindent
2. To guarantee a clear identification of a muon track from $Z^0$ decay in the muon and tracker
systems and determination of its parameters we put the following restrictions on muons
\footnote{Most of the muon selection cuts are taken from \cite{MS,CMJ}.}:
~\\[-5mm]

(a) on the $\Pt$ value of any considered muon:\\[-7pt]
\begin{equation}
\Pt^\mu \geq 10 ~GeV/c;
\label{eq:sc2a}
\end{equation}

\vskip-2.0mm
(b) on the $\Pt$ value of the most energetic muon in a pair:\\[-5pt]
\begin{equation}
\Pt^\mu_{max} \geq \Pt^\mu_{CUT}
\label{eq:sc3b}
\end{equation}
($\Pt^\mu_{CUT}\geq 20~GeV/c$ and depends on the energy scale; see Fig.~6 of Section 4.3);\\[-5mm]

(c) on the value of the ratio of $\Pt^{isol}$, i.e. 
the scalar sum of $\Pt$ of all particles surrounding a muon, to $\Pt^\mu$ ($\Pt^{isol}/\Pt^\mu$) 
in the cone of radius $R=0.3$ and
on the value of maximal $\Pt$ of a charged particle surrounding a muon in this cone:\\[-10pt]
\begin{equation}
\Pt^{isol}/\Pt^\mu \leq 0.10, \quad \Pt^{ch} \leq 2 ~GeV/c.
\label{eq:sc3c}
\end{equation}
The isolated high-$\Pt$ tracks can be reconstructed with a good efficiency (at least $98\%$ over all
pseudorapidity region $|\eta|\lt2.4$; see \cite{MS})
and with generation of a low number of fake and ghost tracks.

\noindent
3. A muon is selected in the region of the muon system acceptance:\\[-10pt]
\begin{equation}
 |\eta^\mu|<2.4.
\end{equation}
\noindent
4. To select muon pairs only from the $Z^0$ decay we limit the value of invariant mass
of a muon pair $M_{inv}^{ll}$ by:
\begin{equation}
|M^Z - M_{inv}^{ll}| \leq 5 ~GeV/c^2.
\label{eq:sc3}
\end{equation}

\noindent  
5. We select the events with the vector $\vec{\Pt}^{Jet}$ being ``back-to-back" to
the vector $\vec{\Pt}^{Z}$ (in the plane transverse to the beam line)
within $\dphi$ defined by the equation:\\[-5pt]
\begin{equation}
\phizj=180^\circ \pm \Delta\phi 
\label{eq:sc4}
\end{equation}
where $\phizj$ is the angle between the $\vec{\Pt}^Z$ and $\vec{\Pt}^{Jet}$ vectors: 
$\vec{\Pt}^{Z}\vec{\Pt}^{Jet}=\Pt^{Z}\Pt^{Jet} cos(\phizj)$, ~~~
where ~$\Pt^{Z}=|\vec{\Pt}^{Z}|,~~\Pt^{Jet}=|\vec{\Pt}^{Jet}|$. 
$\Delta\phi$ defined in the interval $5-15^\circ$ is the most effective choice.

\noindent
6. The initial and final state radiations (ISR and FSR) manifest themselves most clearly
as some final state mini-jets or clusters activity (see the previous section and 
\cite{BKS_P1}--\cite{BKS_P5}). To suppress it, we impose a new cut condition that 
was not formulated in an evident form in previous experiments: we choose the \zpj events
that do not have any other jet-like or cluster high $\Pt$ activity  by taking values of
$\Pt^{clust}$ (the cluster cone $R_{clust}(\eta,\phi)=0.7$), being smaller than some threshold
$\Pt^{clust}_{CUT}$ value, i.e. we select the events with\\[-7pt]
\begin{equation}
\Pt^{clust} \leq \Pt^{clust}_{CUT}.
\label{eq:sc5}
\end{equation}

\noindent
7. We limit the value of the modulus of the vector sum of $\vec{\Pt}$ of all
particles that do not belong to the \zpj system but 
fit into the region $|\eta|\lt5$ covered by
the calorimeter system, i.e., we limit the signal in the cells ``beyond the jet and $Z^0$'' regions
by the following cut:\\[-7pt]
\begin{equation}
\left|\sum_{i\not\in Jet,Z^0}\vec{\Pt}^i\right| \equiv \Pt^{out} \leq \Pt^{out}_{CUT},
~~|\eta|\lt5.
\label{eq:sc6}
\end{equation}

\noindent
The importance of $\Pt^{out}_{CUT}$ and $\Pt^{clust}_{CUT}$
for selection of events with a good balance of \ptzj was already shown in
\cite{BKS_P1} -- \cite{BKS_P5} and in \cite{GPJ_D0}. In Sections 5 and 6 it 
will be demonstrated once again for the case of \zpj events.
The set of selection cuts 1 -- 7 we call as ``Selection 1''.  


\noindent
8. By analogy with \cite{BKS_P1} -- \cite{BKS_P5} and \cite{GPJ_D0} we use 
a ``jet isolation'' requirement (introduced for the first time in \cite{BKS_P1}),
i.e. the presence of a ``clean enough'' (in the sense of limited $\Pt$
activity) region inside the ring of $\Delta R=0.3$ 
around the jet.  Following this picture, we restrict the ratio of the scalar sum
of transverse momenta of particles belonging to this ring, i.e.\\[-10pt]
\begin{equation}
\Pt^{ring}/\Pt^{jet} \equiv \epsilon^{jet}
, \quad {\rm where ~~~~ }
\Pt^{ring}=\!\!\!\sum\limits_{\footnotesize i \in 0.7<R<1} \!|\vec{\Pt}^i|
\label{eq:sc7}
\end{equation}
~\\[-4pt]
with $\epsilon^{jet}\leq 3-8\%$ (see Sections 6 and 7).
The set of events that 
pass cuts 1 -- 7 will be called ``Selection 2''.

\noindent
9. As in \cite{BKS_P1,GPJ_D0} in  the following ``Selection 3'' 
we shall keep only those events in which
one and the same jet (i.e. up to good accuracy having the same values
of $\Pt^{jet}, ~R^{jet}$ and $\Delta\phi$) is found simultaneously by
every of three jetfinders used here: UA1, UA2 and LUCELL.
For these jets (and also clusters) we require the following conditions:\\[-15pt]
\begin{eqnarray}
\Pt^{Jet}\gt30~GeV/c, \qquad \Pt^{clust}\lt \Pt^{clust}_{CUT},\qquad
\dphi\lt15^\circ, \qquad \epsilon^{jet} \leq 5\%.
\label{eq:sc8}
\end{eqnarray}


\noindent 
10. As we shown in \cite{BKS_P1,GPJ_D0} 
One can expect reasonable results of modeling the jet energy calibration procedure
and subsequent practical realization only if one uses a set of selected events with small 
missing transverse momentum $\Pt^{miss}$. We define it here as a $\Pt$ vector sum of 
all the particles flying mostly in the direction of the non-instrumented region $|\eta|>5.0$ 
 and neutrinos with $|\eta|\lt5.0$:\\[-10pt]
\begin{eqnarray}
 \vec{\Pt}^{miss} = \vec{\Pt}^{|\eta|>5.0} + \sum\limits_{i \in |\eta|\lt5.0} \vec{\Pt}_{(\nu)}^i.
\label{eq:miss}
\end{eqnarray}
Here $\vec{\Pt}^{|\eta|\gt5}$ is the total transverse momentum of non-observable particles
$i$ flying in the direction of the non-instrumented forward part of the CMS detector
($|\eta|\gt5$):\\[-8pt]
\begin{equation}
\sum\limits_{i
\in |\eta| \gt 5} \vec{\Pt}^i \equiv \vec{\Pt}^{|\eta|\gt5}.  
\label{eq:eta5}
\end{equation}
We shall use the following cut on ${\Pt}^{miss}$:\\[-15pt]
\begin{eqnarray}
\Pt^{miss}~\leq \Pt^{miss}_{CUT}.
\label{eq:sc11}
\end{eqnarray}
The aim of the event selection with small $\Pt^{miss}$
is quite obvious: we need a set of events with a reduced
$\Pt^{Jet}$ uncertainty due to a possible presence of a non-detectable
neutrino contribution to a jet, for example.

The exact values of the cut parameters $\Pt^{\mu}_{CUT}$,
$\epsilon^{jet}$, $\Pt^{clust}_{CUT}$, $\Pt^{out}_{CUT}$
will be specified below, since they may be different, for instance, 
for various $\Pt^{Z}$ intervals.

\subsection{The $\Pt$-balance equation of \zpj event.}                            

The conservation law for \zpj events as a whole can be written in the following vector 
form \cite{BKS_P1,GPJ_D0}:\\[-12pt]
\begin{eqnarray}
\vec{\Pt}^{Z} +
\vec{\Pt}^{Jet} +
\vec{\Pt}^{O}+
\vec{\Pt}^{|\eta|>5} = 0.
\label{eq:vec_bal}
\end{eqnarray}
$\vec{\Pt}^{|\eta|>5}$ is defined in (\ref{eq:eta5}) and
$\vec{\Pt}^{O}$ is a total transverse momentum of all other ($O$) particles besides 
``jet particles and muons from $Z^0$ decay'' (\zpj system) 
in the $|\eta|\lt5$ region and defined as:\\[-5pt]
\begin{equation}
 \vec{\Pt}^{O} =
\vec{\Pt}^{out}+\vec{\Pt}^{O}_{(\nu)}+\vec{\Pt}^{O}_{(\mu, |\eta^\mu|\gt2.4)}.
\label{eq:pto}
\end{equation}
In its turn, $\vec{\Pt}^{out}$ is a sum of clusters $\Pt$ (with $\Pt^{clust}$ smaller than $\Pt^{Jet}$) 
and $\Pt$ of single hadrons ($h$), photons ($\gamma$) and electrons ($e$) with $|\eta| \lt 5$
and muons ($\mu$) with $|\eta^\mu| \lt 2.4$ that are out of the \zpj system:\\[-5pt]
\begin{equation}
\vec{\Pt}^{out} =
\vec{\Pt}^{clust}
+\vec{\Pt}^{sing}_{(h)}
+\vec{\Pt}^{nondir}_{(\gamma)}
+\vec{\Pt}^{}_{(e)}+\vec{\Pt}^{O}_{(\mu, |\eta^\mu|\lt2.4)}, \quad  |\eta|\lt5.
\label{eq:pt_o}
\end{equation}
The last two terms in equation (\ref{eq:pto}) are the transverse momentum
carried out by the neutrinos that do not belong to the jet but that are
contained in the $|\eta| \lt 5$ region ($\vec{\Pt}^{O}_{(\nu)}$) and non-detectable muons
flying with $|\eta^\mu|\gt2.4$ ($\vec{\Pt}^{O}_{(\mu, |\eta^\mu|\gt2.4)}$).

To conclude this section, let us rewrite
the basic vector $\Pt$-balance equation 
in the following scalar form, more suitable to present the final results: \\[-5pt]
\begin{equation}
\frac{\Pt^{Z}-\Pt^{Jet}}{\Pt^{Z}}=(1-cos\dphi) 
+ \Db/\Pt^{Z}, \label{eq:sc12}
\label{eq:sc_bal}
\end{equation}
~\\[-2pt]
where
$\Db\equiv (\vec{\Pt}^{O}+\vec{\Pt}^{|\eta|\gt5)})\cdot \vec{n}^{Jet}$ ~with
$\vec{n}^{Jet}=\vec{\Pt}^{Jet}/\Pt^{Jet}$ and 
$\dphi$ is the angle that enters equation (\ref{eq:sc4}).

As will be shown in Section 6, the first term on the
right-hand side of equation (\ref{eq:sc_bal}) is negligibly
small and tends to decrease fast with growing $\Pt^{Jet}$. 
So, the main contribution to the $\Pt$ disbalance in the
\zpj system is caused by the term $\Db/\Pt^{Z}$ \cite{BKS_P1}--\cite{BKS_P5}, \cite{GPJ_D0}.

\section{Estimation of a non-detectable part of $\Pt^{jet}$.}       
This subject is considered in detail in \cite{BKS_P1}. Here we outline the main results
for the case of \zpj events.
One of the main sources of this part, that can be estimated on the particle level,
is non-detectable particles (like neutrinos and muons
with $|\eta|\gt2.4$)
\footnote{In a real experiment, of course, it can be also conditioned by 
many other reasons as, for instance, the energy leakage
due to constructive features of the detector, magnetic filed effects and so on.}

The missing transverse momentum $\Pt^{miss}$ (see (\ref{eq:miss}))and a $\Pt$ contribution 
to a jet from non-detectable particles are estimated here in the framework of simulation 
with PYTHIA
\footnote{We have considered the case of switched-off decays of
$\pi^{\pm}$ and $\;K^{\pm}$ mesons (according to the PYTHIA default
agreement, $\pi^{\pm}$ and $\;K^{\pm}$ mesons are stable).
}.
The detailed information about the transverse momenta of non-detectable neutrinos
$\Pt^{Jet}_{(\nu)}$ averaged over all events
(no cut on $\Pt^{miss}$ was used) as well as about mean $\Pt$ values
of muons belonging to jets $\la \Pt^{Jet}_{(\mu)}\ra$ is presented
in Tables 1--12 of Appendix 1 for the sample of events with jets which are
entirely contained in the barrel region of the
calorimeter ($|\eta^{jet}|\lt1.4$, ``HB-events'', see Section 4 and 5).
In these tables the ratio of number of the events with non-zero $\Pt^{Jet}_{(\nu)}$
to the total number of events is denoted by $R^{\nu \in Jet}_{event}$ and
the ratio of the number of events with non-zero $\Pt^{Jet}_{(\mu)}$
to the total number of events is denoted by $R^{\mu \in Jet}_{event}$.
%
%

A wide variation of $\Pt^{miss}_{CUT}$ as well as the case of allowed $K^{\pm}$ decays
in the calorimeter volume were presented in detail in \cite{BKS_P1}.
We choose here for the following analysis $\Pt^{miss}_{CUT}=10 ~GeV/c$ found to be optimal
in  \cite{BKS_P1}.

\section{Event rates for different $\Ptz$ and $\eta^Z$
 intervals.}

\subsection{Dependence of the distribution of the number of events
on the "back-to-back" angle $\phizj$ and on $\Pt^{ISR}$. }            
%

Here we study the spectrum of the variable $\Pt{56}$ for the sample of signal events
\footnote{$\Pt{56}$ is approximately proportional to $\Pt^{ISR}$ up to the value of intrinsic parton
transverse momentum $k_T$ inside a proton 
($\la k_T\ra$ was taken to be fixed at the PYTHIA default value, i.e. $\la k_T\ra=0.44\,GeV/c$).}.
For this aim four samples of \zpj events (each by $5\cdot 10^6$) were
generated  by using PYTHIA with 2 subprocesses (2a) and (2b)
and with minimal $\Pt$ of hard $2\to 2$ scattering 
\footnote{CKIN(3) parameter in PYTHIA}
$\pth=20,35,50,75~GeV/c$ 
to cover four $\Pt^{Z}$ intervals: 40--50, 70--85, 100--120, 150--200 $GeV/c$, respectively.
The obtained cross sections for these subprocesses are given in Table ~\ref{tab:cross4}.
~\\[-20pt]
\begin{table}[h]
\begin{center}
\caption{The cross sections (in $microbarns$) of the $qg\to q+Z^0$ and $q\overline{q}\to g+Z^0$ subprocesses
for four $\pth$ values.}
\normalsize
\vskip.1cm
\begin{tabular}{||c||c|c|c|c|}                  \hline \hline
\label{tab:cross4}
Subprocess& \multicolumn{4}{c|}{ $\pth$ values ($GeV/c$)} \\\cline{2-5}
   type   & 20 & 35 & 50  & 75 \\\hline \hline
$qg\to q+Z^0$           & 3.83$\cdot10^{-4}$& 1.71$\cdot10^{-4}$& 9.14$\cdot10^{-5}$& 3.80$\cdot10^{-5}$ \\\hline
$q\overline{q}\to g+Z^0$& 1.20$\cdot10^{-4}$& 0.42$\cdot10^{-4}$& 1.93$\cdot10^{-5}$& 0.69$\cdot10^{-5}$ \\\hline
Total                 & 5.03$\cdot10^{-4}$& 2.13$\cdot10^{-4}$& 1.11$\cdot10^{-4}$& 4.59$\cdot10^{-5}$ \\\hline
\end{tabular}
\end{center}
\vskip-4mm
\end{table}

~\\[-14mm]

For our analysis  we used cuts  (\ref{eq:sc1}) -- (\ref{eq:sc6}) 
and the following cut parameters: \\[-12pt]
\begin{eqnarray}
\Pt^{\mu}_{max}\gt20\;GeV/c, \quad
\dphi\lt15^{\circ}, \quad
\Pt^{clust}_{CUT}=30\;GeV/c.
\label{eq:gen_cuts}
\end{eqnarray}
~\\[-11mm]

In Tables \ref{tab:pt56-1}, \ref{tab:pt56-2} and \ref{tab:pt56-4},
\ref{tab:pt56-5}  we study (as in \cite{BKS_P1}) $\Pt56$ spectra for
two most illustrative cases of $\Pt^{Z}$ intervals $40\lt\Pt^{Z}\lt50 ~GeV/c$ (Tables 2 and 5) and
$100\lt\Pt^{Z}\lt120 ~GeV/c$ (Tables 3 and 6). The distributions of the
 number of events for the integrated luminosity $L_{int}=10\,fb^{-1}$
in different $\Pt56$ intervals  and for different ``back-to-back'' angle intervals
$\phizj=180^\circ \pm \dphi~$ (with $\dphi=15^\circ,\,10^\circ$ and $5^\circ$
as well as without any restriction on $\dphi$, i.e. for the whole $\phi$ interval
$\Delta\phi=180^\circ$)
 are given there. The LUCELL jetfinder was used to find jets and clusters.
Tables \ref{tab:pt56-1} and \ref{tab:pt56-2} correspond to the events selected with
cuts $\Pt^{clust}\lt30\,GeV/c$ and
without any limit on $\Pt^{out}$ value,
while Tables~\ref{tab:pt56-4} and \ref{tab:pt56-5}  correspond to more
restrictive selection cuts $\Pt^{clust}\lt10\,GeV/c$ and  $\Pt^{out}\lt10\,GeV/c$.

First, from the last summary lines of Tables
\ref{tab:pt56-1}, \ref{tab:pt56-2} and \ref{tab:pt56-4},
\ref{tab:pt56-5} we can make a general conclusion about the $\dphi$ dependence
of the event spectrum.
In the case when no restriction is used   
we can see that for the $40\leq \Pt^{Z}\leq 50 ~GeV/c$ 
%
\def\baselinestretch{0.98}
\begin{table}[htbp]
\begin{center}
\vskip-1.2cm
\caption{Number of events dependence on $\Pt56$ and
$\Delta\phi$ for $40\leq \Pt^{Z}\leq 50 \, GeV/c$}
\vskip-3pt
{\footnotesize and $\Pt^{clust}_{CUT}= 30 \, GeV/c$~ for $L_{int}$=10$\,fb^{-1}$.}
\vskip.2cm
\begin{tabular}{||c||r|r|r|r||} \hline \hline
\label{tab:pt56-1}
 $\Pt{56}$&\multicolumn{4}{|c||}{ $\dphi_{max}$} \\\cline{2-5}
 $(GeV/c)$  &\aaa $180^\circ$\aaa&\aaa$15^\circ$\aaa&\aaa$10^\circ$\aaa&\aaa$5^\circ$\aaa \\\hline\hline
    0 --   5 &                  18525  &    16965 &     15880 &     12708\\\hline
    5 --  10 &                  29094  &    26671 &     23419 &     13579\\\hline
   10 --  15 &                  24192  &    19935 &     14042 &      7033\\\hline
   15 --  20 &                  18168  &    10910 &      7088 &      3481\\\hline
   20 --  25 &                  13424  &     5833 &      3924 &      1968\\\hline
   25 --  30 &                  10169  &     3604 &      2380 &      1172\\\hline
   30 --  40 &                  14070  &     4114 &      2677 &      1311\\\hline
   40 --  50 &                   7544  &     1833 &      1184 &       618\\\hline
   50 -- 100 &                   5904  &     1727 &      1097 &       550\\\hline
  100 -- 300 &                      8  &        3 &         2 &         0\\\hline
  300 -- 500 &                      0  &        0 &         0 &         0\\\hline
   30 -- 500 &                 141095  &    91594 &     71694 &     42423\\\hline
\end{tabular}
\vskip0.2cm
\caption{Number of events dependence on $\Pt56$ and
$\Delta\phi$ for $100\leq \Pt^{Z}\leq 120 \, GeV/c$}
\vskip-3pt
{\footnotesize  and $\Pt^{clust}_{CUT}= 30 \, GeV/c$ for $L_{int}$=10$\,fb^{-1}$.}
\vskip0.2cm
\begin{tabular}{||c||r|r|r|r||} \hline \hline
\label{tab:pt56-2}
 $\Pt{56}$  &\multicolumn{4}{c||}{ $\dphi_{max}$} \\\cline{2-5}
 $(GeV/c)$  &\aaa $180^\circ$\aaa&\aaa$15^\circ$\aaa&\aaa$10^\circ$\aaa&\aaa$5^\circ$\aaa \\\hline\hline
    0 --   5 &                   1849 &      1837  &     1790   &    1616 \\\hline
    5 --  10 &                   3798 &      3770  &     3667   &    3247 \\\hline
   10 --  15 &                   3635 &      3600  &     3477   &    2542 \\\hline
   15 --  20 &                   3065 &      3025  &     2847   &    1592 \\\hline
   20 --  25 &                   2491 &      2424  &     1976   &     986 \\\hline
   25 --  30 &                   2115 &      2000  &     1418   &     709 \\\hline
   30 --  40 &                   2507 &      2039  &     1398   &     721 \\\hline
   40 --  50 &                   1061 &       744  &      527   &     289 \\\hline
   50 -- 100 &                   1105 &       768  &      582   &     325 \\\hline
  100 -- 300 &                    194 &       147  &      107   &      63 \\\hline
  300 -- 500 &                      2 &         2  &        1   &       0 \\\hline
    0 -- 500 &                  21826 &     20356   &   17797     & 12094 \\\hline
\end{tabular}
\vskip0.2cm
\caption{Number of events dependence on $\dphi$ and on
$\Pt^{Z}$ for $L_{int}=10\,fb^{-1}$.}
\vskip-3pt
{\footnotesize $\Pt^{clust}_{CUT}=30 ~GeV/c$ (summary).}
\vskip0.2cm
\begin{tabular}{||c||r|r|r|r||} \hline \hline
\label{tab:pt56-3}
 $\Pt^{Z}$  &\multicolumn{4}{c||}{ $\dphi_{max}$} \\\cline{2-5}
 $(GeV/c)$  &\aaa $180^\circ$\aaa&\aaa$15^\circ$\aaa&\aaa$10^\circ$\aaa&\aaa$5^\circ$\aaa \\\hline\hline
 40 -- 50  &    141095 &     91591  &    71694  &    42423  \\\hline 
 70 -- 80  &     40032 &     32551  &    26710  &    16794  \\\hline  
100 -- 120 &      2182 &     20356  &    17797  &    12094  \\\hline 
150 -- 200 &      8649 &      8558  &     8134  &     6182  \\\hline  
\end{tabular}
\end{center}
\end{table}

\begin{table}[htbp]
\begin{center}
\caption{Number of events dependence on $\Pt56$ and
$\Delta\phi$ for $40\leq \Pt^{Z}\leq 50 \, GeV/c$}
\vskip-3pt
{\footnotesize  and $\Pt^{clust}_{CUT}= 10~  GeV/c$ and $\Pt^{out}_{CUT}= 10~ GeV/c$
for $L_{int}$=10$\,fb^{-1}$.}
\vskip0.2cm
\begin{tabular}{||c||r|r|r|r||} \hline \hline
\label{tab:pt56-4}
 $\Pt{56}$  &\multicolumn{4}{c||}{ $\dphi_{max}$} \\\cline{2-5}
 $(GeV/c)$  &\aaa $180^\circ$\aaa&\aaa$15^\circ$\aaa&\aaa$10^\circ$\aaa&\aaa$5^\circ$\aaa \\\hline\hline
    0 --   5 &    11619  &    11603   &   11409   &    9603  \\\hline
    5 --  10 &    15329  &    15258   &   14288   &    8767  \\\hline
   10 --  15 &     6787  &     6479   &    5156   &    2768 \\\hline
   15 --  20 &     1810  &     1533   &    1204   &     645 \\\hline
   20 --  25 &      677  &      527   &     432   &     253  \\\hline
   25 --  30 &      305  &      238   &     195   &     119 \\\hline
   30 --  40 &      277  &      222   &     193   &     111  \\\hline
   40 --  50 &      127  &      111   &      91   &      44  \\\hline
   50 -- 100 &       36  &       32   &      24   &      12 \\\hline
  100 -- 300 &        0  &        0   &       0   &       0 \\\hline
  300 -- 500 &        0  &        0   &       0   &       0  \\\hline
    0 -- 500 &    36967  &    35996   &   32987   &   22315 \\\hline
\end{tabular}
\vskip0.2cm
\caption{Number of events dependence on $\Pt56$ and
$\Delta\phi$ for $100\leq \Pt^{Z}\leq 120 \, GeV/c$ }
\vskip-3pt
{\footnotesize and $\Pt^{clust}_{CUT}= 10~  GeV/c$ and $\Pt^{out}_{CUT}= 10~ GeV/c$
for $L_{int}$=10$\,fb^{-1}$.}
\vskip0.2cm
\begin{tabular}{||c||r|r|r|r||} \hline \hline
\label{tab:pt56-5}
 $\Pt{56}$  &\multicolumn{4}{c||}{ $\dphi_{max}$} \\\cline{2-5}
 $(GeV/c)$  &\aaa $180^\circ$\aaa&\aaa$15^\circ$\aaa&\aaa$10^\circ$\aaa&\aaa$5^\circ$\aaa \\\hline\hline
    0 --   5 &     1133   &    1133   &    1133   &    1121   \\\hline
    5 --  10 &     1932   &    1932   &    1932   &    1877  \\\hline
   10 --  15 &     1002   &    1002   &    1002   &     867  \\\hline
   15 --  20 &      309   &     309   &     309   &     234  \\\hline
   20 --  25 &       95   &      95   &      91   &      63   \\\hline
   25 --  30 &       49   &      49   &      45   &      33 \\\hline
   30 --  40 &       48   &      44   &      40   &      32  \\\hline
   40 --  50 &       27   &      25   &      25   &      25 \\\hline
   50 -- 100 &       44   &      44   &      44   &      40  \\\hline
  100 -- 300 &        5   &       5   &       5   &       5  \\\hline
  300 -- 500 &        0   &       0   &       0   &       0  \\\hline
    0 -- 500 &     4641   &    4637   &    4621   &    4293  \\\hline
\end{tabular}
\vskip0.2cm
\caption{Number of events dependence on $\dphi$ and on $\Pt^{Z}$ for $L_{int}=10\,fb^{-1}$.}
\vskip-3pt
{\footnotesize $\Pt^{clust}_{CUT}= 10~  GeV/c$ and $\Pt^{out}_{CUT}= 10~ GeV/c$ (summary).}
\vskip0.2cm
\begin{tabular}{||c||r|r|r|r||} \hline \hline
\label{tab:pt56-6}
 $\Pt^{Z}$  &\multicolumn{4}{c||}{ $\dphi_{max}$} \\\cline{2-5}
 $(GeV/c)$  &\aaa $180^\circ$\aaa&\aaa$15^\circ$\aaa&\aaa$10^\circ$\aaa&\aaa$5^\circ$\aaa \\\hline\hline
 40 -- 50  &     36967  &    35996  &    32987  &    22315  \\\hline   
 70 -- 80  &      8688  &     8657  &     8542  &     7033   \\\hline      
100 -- 120 &      4641  &     4637  &     4621  &     4293 \\\hline       
150 -- 200 &      1746  &     1746  &     1742  &     1719  \\\hline     
\end{tabular}
\end{center}
\end{table}

\def\baselinestretch{1.0}

\noindent
(Table \ref{tab:pt56-1})
interval about 65$\%$ of events are concentrated in the $\Delta\phi\lt15^\circ$ range, 
while 30$\%$ of events are in the $\Delta\phi\lt5^\circ$ range. At the same time the analogous 
summary line of Table \ref{tab:pt56-2}
shows us that for $100\leq \Pt^{Z}\leq 120\, GeV/c$ the event spectrum
moves noticeably to the small $\dphi$ region: more than 94$\%$ of events have
$\Delta\phi\lt15^\circ$ and 56$\%$ of them have $\Delta\phi\lt5^\circ$.

We observe a tendency of the distributions of the number of signal \zpj events to be
concentrated in a rather narrow back-to-back angle interval 
$\Delta\phi\lt15^\circ$ with $\Ptz$ growing. It becomes more distinct with a more restrictive cuts
$\Pt^{out}_{CUT}= 10\,GeV/c$ and $\Pt^{out}_{CUT}= 10\,GeV/c$ 
(Tables \ref{tab:pt56-4} and
\ref{tab:pt56-5}). From the last summary line of Table \ref{tab:pt56-4} we see for these cuts
that in the case of $40\leq \Pt^{Z}\leq 50\, GeV/c~$ more than
$96\%$ of the events have $\Delta\phi\lt15^\circ$, while $60\%$ of them are
in the $\Delta\phi\lt5^\circ$ range. For $100\leq \Pt^{Z}\leq 120\, GeV/c$
(see Table \ref{tab:pt56-5}) more than  $92\%$ of the events, subject to these cuts,
have $\Delta\phi\lt5^\circ$. It means that while suppressing $\Pt$ activity beyond the \zpj system
by imposing $\Pt^{clust}_{CUT}= 10 ~GeV/c$ and $\Pt^{out}_{CUT}= 10\,GeV/c$
we can select the sample of events with a clean
back-to-back ($\Delta\phi\lt15^\circ$) topology of $\vec{\Pt}^Z$ and $\vec{\Pt}^{jet}$ orientation
%
\footnote{An increase in \ptz produces the same effect, as is seen
from Tables \ref{tab:pt56-2} and \ref{tab:pt56-4}, and
will be demonstrated in more detail in Section 6 and Appendices 2--5.}.

The other lines of Tables \ref{tab:pt56-1}, \ref{tab:pt56-2} and
\ref{tab:pt56-4}, \ref{tab:pt56-5} contain the information
about the $\Pt56$ spectrum (or, up to $k_T$ effect, $\Pt^{ISR}$ spectrum).

From the comparison of Table \ref{tab:pt56-1}  with  Table \ref{tab:pt56-4}
(as well as from Tables \ref{tab:pt56-2} and \ref{tab:pt56-5})
one can conclude that the width of the most populated part of the $\Pt56$ (or $\Pt^{ISR}$)
spectrum is noticeably reduced with restricting $\Pt^{clust}_{CUT}$ and $\Pt^{out}_{CUT}$. 

We supply Tables \ref{tab:pt56-1}, \ref{tab:pt56-2} and \ref{tab:pt56-4}, \ref{tab:pt56-5} with 
summarizing Tables \ref{tab:pt56-3} and \ref{tab:pt56-6}
containing an illustrative information on $\Delta\phi$ dependence of the total number of events.
They include more $\Pt^{Z}$ intervals and contain analogous numbers of events that can be collected
in different $\Delta\phi$ intervals for 
$\Pt^{clust}_{CUT}$, $\Pt^{out}_{CUT}$ and other cuts, defined by (\ref{eq:gen_cuts}), at $L_{int}=10\,fb^{-1}$.

We can conclude from Tables \ref{tab:pt56-1}--\ref{tab:pt56-6} 
that restriction on the $\Pt^{clust}_{CUT}$ and $\Pt^{out}_{CUT}$ variables are
good tools to reduce ISR while by limiting $\dphi$ angle the ISR remains, in fact, without a change.
Meanwhile, in spite of about twofold spectra reduction of the ISR (or $\Pt56$), see Tables 4 and 7,
it continues to be noticeable at the LHC energies
\footnote{The analogous conclusion was done by studying \gpj events in \cite{BKS_P1}.}.
%

\subsection{$\Ptz$, $\eta^{Z}$ and $\Pt^\mu$ dependence of rates.}
~\\[-30pt]
\begin{flushleft}
\parbox[r]{.48\linewidth}{
In Table \ref{tab:pt-eta} we present the number of events calculated after passing selection cuts
(\ref{eq:sc1})--(\ref{eq:sc6}) for different $\Ptz$ and $\eta^{Z}$ intervals (lines and columns of
the table, respectively).
The last column of this table contains the total number
of events (at $L_{int}=10\,fb^{-1}$) at
$|\eta^Z|\lt5.0$ for a given $\Ptz$ interval.
We see that the number of events  decreases fast
with growing $\Ptz$
(but it decreases much slower as compared with decrease in \ptg ~spectrum
in the case of \gpj events, see \cite{BKS_P1}).
 It also drops with growing $|\eta^Z|$ starting from $|\eta^Z|\approx 2.0$
and has weak dependence on $\eta^Z$ in the interval $|\eta^Z|<2.0$.
~The~ analogous~ information is illustrated by Fig. 5 for  three
 $\Ptz$ intervals
}
\end{flushleft}
\begin{flushright}
\begin{figure}[htbp]
 \vspace{-73mm}
 \hspace*{83mm} \includegraphics[width=100mm,height=7.5cm]{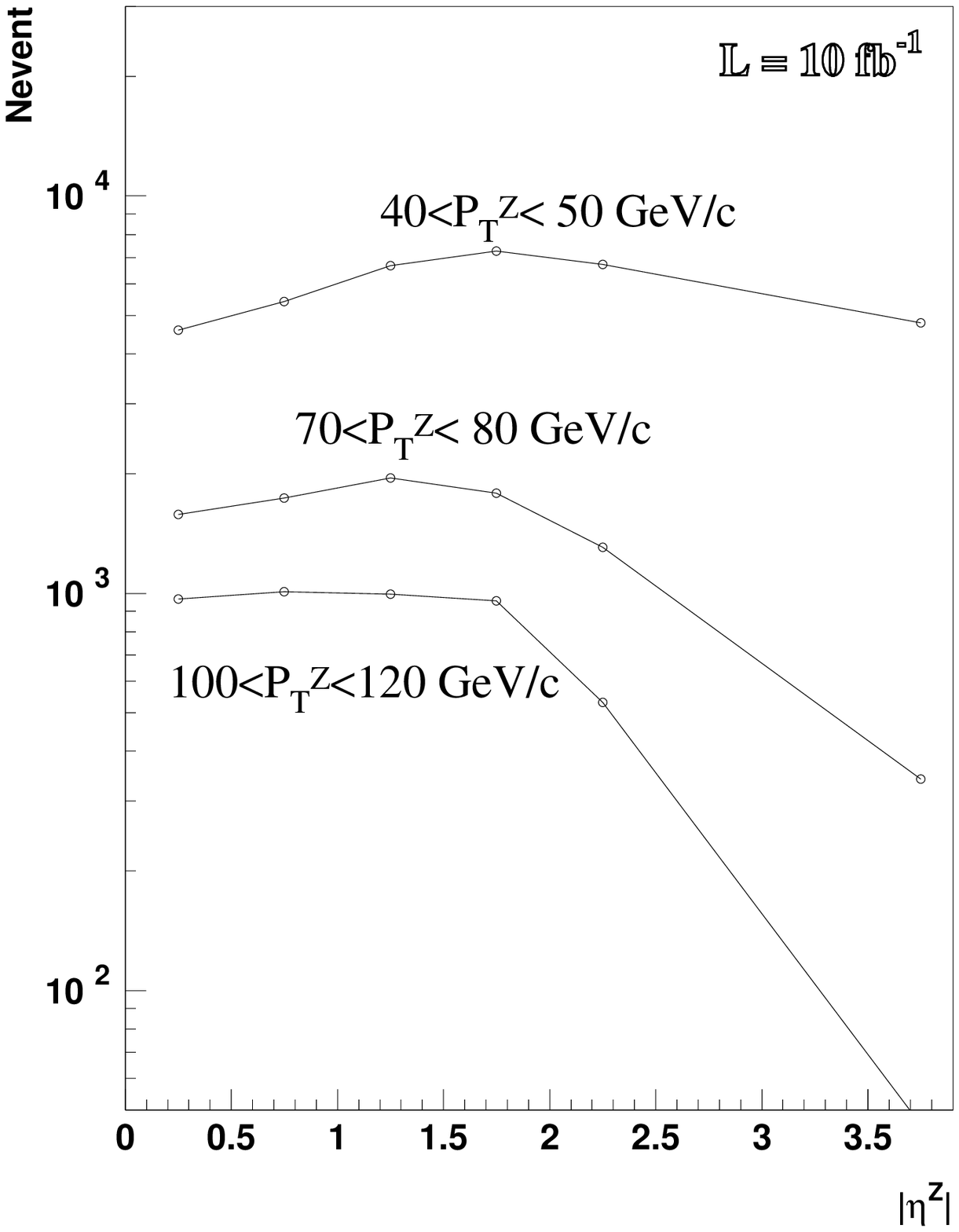}%
\end{figure}
\end{flushright}
\vspace{-8mm}
\hspace*{89mm} {\footnotesize {Fig.~5: $\eta$-dependence of rates for
different $\Ptz$ intervals.\\}}
~\\[-21mm]
\hspace*{60mm}~\footnote{We have limited $Z^0$ pseudorapidity spectrum from above 
in Fig.~5 and Table \ref{tab:pt-eta} only 
to give understanding about the its behavior inside this $\eta^Z$ interval and, 
certainly, have not used those limits as cuts anywhere in this paper.}.
~\\[19mm]
~\\[-10mm]
\def\baselinestretch{1.0}
\begin{table}[h]
\begin{center}
\caption{Rates for $L_{int}=10\,fb^{-1}$ for different intervals of $\Pt^{Z}$ and 
$\eta^{Z}$ ($\Pt^{clust}_{CUT}= 10 \, GeV/c$, ~$\Pt^{out}_{CUT}= 10~ GeV/c$  
and $\Delta\phi \leq 15^\circ$).}
\vskip0.2cm
\begin{tabular}{||c||r|r|r|r|r|r||r||} \hline \hline
\label{tab:pt-eta}
$\Pt^{Z}$ &\multicolumn{6}{c||}{$|\Delta \eta^{Z}|$~~ intervals}
&all ~~$|\eta^{Z}|$ \\\cline{2-8}
$(GeV/c)$ & 0.0-0.5 & 0.5-1.0 & 1.0-1.5 & 1.5-2.0 & 2.0-2.5 & 2.5-5.0 & 0.0-5.0
 \\\hline \hline
 40 -- 50 &     4594 &    5425  &   6673 &    7267 &    6732 &    4796  &  35486\\\hline
 50 -- 60 &     3128 &    3509  &   4297 &    4570 &    3976 &    2000  &  21471\\\hline
 60 -- 70 &     2253 &    2443  &   2855 &    2934 &    2229 &     851  &  13567\\\hline
 70 -- 80 &     1580 &    1734  &   1948 &    1786 &    1307 &     341  &   8692\\\hline
 80 -- 90 &     1152 &    1148  &   1267 &    1236 &     824 &     170  &   5790\\\hline
 90 --100 &      741 &     859  &    812 &     808 &     523 &      59  &   3802\\\hline
100 --110 &      582 &     590  &    594 &     546 &     305 &      36  &   2657\\\hline
110 --120 &      384 &     428  &    451 &     412 &     226 &       8  &   1905\\\hline
120 --140 &      523 &     582  &    562 &     531 &     293 &      12  &   2503\\\hline
140 --170 &      392 &     380  &    368 &     341 &     190 &       4  &   1675\\\hline
170 --200 &      170 &     186  &    162 &     170 &      63 &       2  &    756\\\hline
200 --240 &      111 &     103  &     99 &      91 &      40 &       0  &    444\\\hline
240 --300 &       71 &      51  &     44 &      48 &      20 &       0  &    238\\\hline    
\hline
\end{tabular}
\end{center}
\end{table}
\normalsize

~\\[-14mm]

\setcounter{figure}{5}
In Fig.~\ref{fig:pt_mu} we have plotted a normalized distributions of the number of events
over $\Pt$ of muons  from $Z^0$ decay for two $\Ptz$ intervals: $40\lt\Ptz\lt50$ and 
$100\lt\Ptz\lt120 ~GeV/c$. The muon spectra are limited by the condition (4) $\Pt^\mu\gt10~GeV/c$.
We also see that the spectra with muons having maximal $\Pt$ in the pair starts at $20~GeV/c$ for
$40\lt\Ptz\lt50~GeV/c$ and at $50~GeV/c$ for $100\lt\Ptz\lt120 ~GeV/c$. It explains our choice in (5)
for $\Pt^\mu_{max}$ restriction.
\begin{figure}[h]
\vskip-13mm
  \hspace{0mm} \includegraphics[width=16cm,height=7.8cm]{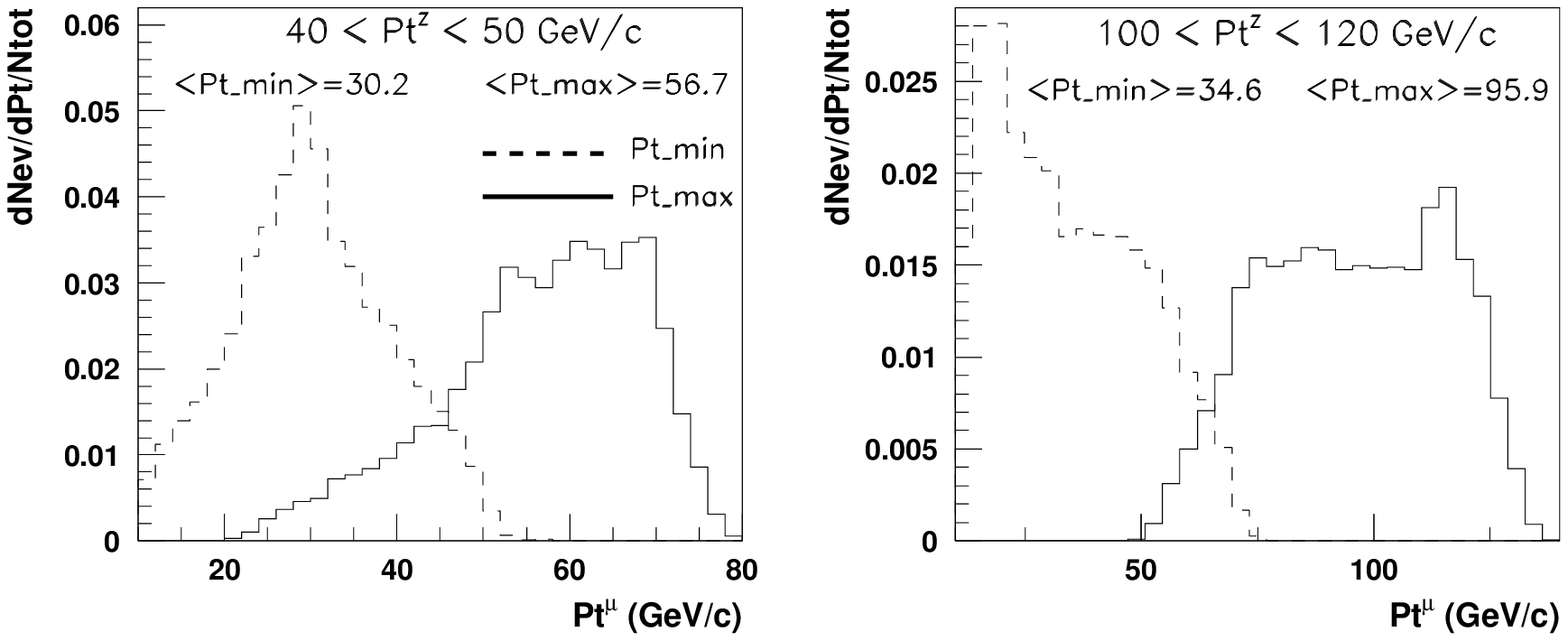}
\vskip-8mm
   \caption{A normalized distributions of the number of events over $\Pt$ of muons from $Z^0$ decay:
for a muon with  maximal $\Pt$ (full line) and for a muon with minimal $\Pt$ (dashed line)
in the pair.}
\vskip 7mm
\label{fig:pt_mu}
\end{figure}

\subsection{Estimation of \zpj event rates for the HB, HE and HF regions.} 

Since a jet is a wide-spread object, we present the $\eta^{jet}$ dependence of rates
(for different $\Ptz$ intervals) in a different way.
Namely, Tables \ref{tab:sh1}--\ref{tab:sh4}
 include the rates of events (at $L_{int}=10\,fb^{-1}$) for different
$\eta^{jet}$ intervals, covered by the Barrel, Endcap and Forward
(HB, HE and HF) parts 
of the calorimeter. 
The events are selected after the cuts  (\ref{eq:sc1}) -- (\ref{eq:sc6}) (Selection 1)
with the following values of the cut parameters: \\[-0pt]
\begin{equation}
\Delta \phi\lt15^{\circ}, \quad \Pt^{clust}_{CUT}=10\;GeV/c, \quad \Pt^{out}_{CUT}=10\;GeV/c.
\end{equation}
The first columns of these tables give the
number of events with jets (found by the LUCELL jetfinding algorithm of PYTHIA),
all particles of
 which are comprised entirely (100$\%$) in the Barrel part (HB) and there is a
$0\%$ sharing ($\Delta\Pt^{jet}\!=\!0$) of $\Pt^{jet}$ between the HB and
the neighboring HE part of the calorimeter. The second columns of the 
tables contain the number of events in which $\Pt$ of the jet is shared
between the HB and HE regions. The same sequence of restriction
conditions takes place in the next columns. Thus, the 
HE and HF columns include the number of events with jets
entirely contained in these regions, while the HE+HF  column gives the number of  events where
the jet covers both the HE and HF regions.
From these tables we can see what number of events
can, in principle, be suitable for the most precise jet energy calibration procedure, carried out
 separately for the HB, HE and HF parts of the calorimeter in
different $\Ptz$($\approx\Pt^{jet}$) intervals.
Less restrictive conditions, when up to $10\%$ of the jet $\Pt$
are allowed to be shared between the HB, HE and HF parts of the calorimeter, are given in 
Tables \ref{tab:sh2} and \ref{tab:sh4}. Tables \ref{tab:sh1}  and \ref{tab:sh2}
correspond to the case of Selection 1.
Tables \ref{tab:sh3} and \ref{tab:sh4} contain the number of events collected
with the added Selection 2 restriction (with $\epsilon^{jet}\lt5\%$), i.e.
they include only the events with ``isolated jets'' (defined in Section 2.2).
The reduction factor of about 2 for the  number of  events can be 
found by comparing Tables \ref{tab:sh1} and \ref{tab:sh2} with Tables \ref{tab:sh3} and \ref{tab:sh4}.

From the last summarizing line of Table \ref{tab:sh1} we see 
that for the whole interval $40\lt Pt^{Z}\lt300\; GeV/c$~ PYTHIA predicts
about 45 000 events for HB, 16 000  events for HE  and about 2 000 events for HF
at $L_{int}$=10$\,fb^{-1}$.

~\\[-2pt]
\begin{table}[htpb]
\begin{center}
\caption{Selection 1. $\Delta \Pt^{jet} / \Pt^{jet} = 0.00$ ~($L_{int}$=10$\,fb^{-1}$).}
\vskip0.2cm
\begin{tabular}{||c||c|c|c|c|c||} \hline \hline
\label{tab:sh1}
$\Pt^{Z}$   &     HB  &  HB+HE &     HE  &   HE+HF &     HF\\\hline \hline
 40 --  50  &  15072 &   11179 &    5417 &    3045 &     729 \\\hline
 50 --  60  &   9076 &    7037 &    3231 &    1734 &     376  \\\hline
 60 --  70  &   5813 &    4447 &    2055 &    1030 &     218 \\\hline
 70 --  80  &   3726 &    2903 &    1275 &     669 &     123 \\\hline
 80 --  90  &   2542 &    1901 &     847 &     432 &      67\\\hline
 90 -- 100  &   1711 &    1243 &     558 &     246 &      44 \\\hline
100 -- 110  &   1263 &     879 &     352 &     150 &      12  \\\hline
110 -- 120  &    836 &     681 &     289 &     107 &      20 \\\hline
120 -- 140  &   1085 &     836 &     400 &     154 &       8   \\\hline
140 -- 170  &    752 &     626 &     218 &      71 &       8 \\\hline
170 -- 200  &    348 &     261 &     103 &      44 &       0 \\\hline
200 -- 240  &    206 &     139 &      75 &      20 &       0 \\\hline
240 -- 300  &    111 &      95 &      28 &       4 &       0  \\\hline
 40 -- 300  &  44554 &   34076 &   15789 &    8510 &    2020 \\\hline
\end{tabular}
\vskip1.3cm
\caption{Selection 1. $\Delta \Pt^{jet} / \Pt^{jet} \leq 0.10$ ~($L_{int}$=10$\,fb^{-1}$).}
\vskip0.2cm
\begin{tabular}{||c||c|c|c|c|c||} \hline \hline
\label{tab:sh2}
$\Pt^{Z}$  &     HB   &   HB+HE &     HE  &   HE+HF &     HF\\\hline \hline
 40 --  50 &   19610 &    3251 &   10328  &    887  &   1366  \\\hline
 50 --  60 &   12161 &    1667 &    6439  &    420  &    768\\\hline
 60 --  70 &    7797 &     950 &    4166  &    202  &    444  \\\hline
 70 --  80 &    5077 &     570 &    2633  &    162  &    253  \\\hline
 80 --  90 &    3453 &     372 &    1734  &     83  &    147 \\\hline
 90 -- 100 &    2261 &     242 &    1152  &     48  &     95  \\\hline
100 -- 110 &    1683 &     170 &     729  &     32  &     40\\\hline
110 -- 120 &    1176 &      87 &     582  &     16  &     45\\\hline
120 -- 140 &    1465 &     139 &     816  &     36  &     43\\\hline
140 -- 170 &    1026 &     115 &     511  &     12  &     12\\\hline
170 -- 200 &     475 &      48 &     222  &      5  &      8 \\\hline
200 -- 240 &     273 &      17 &     147  &      3  &      4 \\\hline
240 -- 300 &     158 &      15 &      59  &      0  &      0 \\\hline
 40 -- 300 &   59392 &    8169 &   31395  &   2127  &   3861 \\\hline
\end{tabular}
\end{center}
\end{table}

\newpage
An additional information on the numbers of \zpj events with jets produced by $c$ and $b$ quarks
(see also \cite{BKS_P1} and \cite{MD1,MD2}), given for the integrated luminosity
$L_{int}=10~ fb^{-1}$ for different $\Ptz$($\approx \Pt^{Jet}$) intervals
45--55, 70--85, 100--120 and 150--200 $GeV/c$ is contained in Tables 1--12 of Appendix 1
(they denoted as $Nevent_{(c)}$ and $Nevent_{(b)}$ there).
They also show the ratio  of the number of events caused by gluonic 
Compton-like subprocess (2a) to the number of events due to the sum of
subprocesses (2a) and (2b) ($30sub/all$) and averaged jet radii $\lt\!\!R^{jet}_{gc}\!\!\gt$.
\begin{table}[htpb]
\begin{center}
\vskip0.5cm
\caption{Selection 2. $\Delta \Pt^{jet} / \Pt^{jet} = 0.00$ ~($L_{int}$=10$\,fb^{-1}$).}
\vskip0.2cm
\begin{tabular}{||c||c|c|c|c|c||} \hline \hline
\label{tab:sh3}
$\Pt^{Z}$&        HB  &  HB+HE &     HE  &   HE+HF &     HF\\\hline \hline
 40 --  50 &   6039  &   4221  &   2364  &   1152  &    352\\\hline
 50 --  60 &   4578  &   3398  &   1810  &    847  &    182  \\\hline
 60 --  70 &   3461  &   2637  &   1319  &    645  &    154 \\\hline
 70 --  80 &   2542  &   2020  &    915  &    447  &     91\\\hline
 80 --  90 &   1936  &   1382  &    681  &    329  &     55\\\hline
 90 -- 100 &   1390  &    962  &    475  &    190  &     36\\\hline
100 -- 110 &   1093  &    717  &    305  &    123  &     13 \\\hline
110 -- 120 &    744  &    614  &    273  &     79  &     15\\\hline
120 -- 140 &    990  &    760  &    376  &    158  &      9 \\\hline
140 -- 170 &    713  &    602  &    210  &     71  &      7 \\\hline
170 -- 200 &    341  &    257  &    103  &     45  &      1  \\\hline
200 -- 240 &    206  &    131  &     75  &     19  &      0 \\\hline
240 -- 300 &    111  &     95  &     28  &      4  &      0 \\\hline
 40 -- 300&   24912  &  18489  &   9393  &   4499  &   1169 \\\hline
\end{tabular}
\vskip1.3cm
\caption{Selection 2. $\Delta \Pt^{jet} / \Pt^{jet} \leq 0.10$ ~($L_{int}$=10$\,fb^{-1}$).}
\vskip0.2cm
\begin{tabular}{||c||c|c|c|c|c||} \hline \hline
\label{tab:sh4}
$\Pt^{Z}$  &     HB   &   HB+HE &     HE  &   HE+HF &     HF\\\hline \hline
 40 --  50 &    7770  &   1148 &    4297  &    309  &    602 \\\hline
 50 --  60 &    6083  &    729 &    3425  &    190  &    384 \\\hline
 60 --  70 &    4629  &    554 &    2602  &    119  &    305\\\hline
 70 --  80 &    3465  &    388 &    1885  &     99  &    178  \\\hline
 80 --  90 &    2610  &    249 &    1350  &     59  &    115  \\\hline
 90 -- 100 &    1806  &    190 &     950  &     37  &     79 \\\hline
100 -- 110 &    1434  &    139 &     610  &     23  &     36\\\hline
110 -- 120 &    1057  &     85 &     635  &     31  &     40\\\hline
120 -- 140 &    1338  &    117 &     656  &     21  &     37 \\\hline
140 -- 170 &     974  &    111 &     491  &     12  &     11 \\\hline
170 -- 200 &     467  &     48 &     218  &      4  &      9 \\\hline
200 -- 240 &     273  &     18 &     143  &      4  &      3 \\\hline
240 -- 300 &     158  &     14 &      59  &      0  &      0\\\hline
 40 -- 300 &   33117  &   3952 &   18224  &    990  &   2174 \\\hline
\end{tabular}
\end{center}
\end{table}

\newpage
\section{Features of ~\zpj events in the Barrel region.}
\subsection{Influence of the $\Pt^{clust}_{cut}~$ parameter
on the balance between $Z^0$ and jet transverse momenta
and on the initial state radiation suppression.}

Here we shall study a correlation of $\Pt^{clust}$ with $\Pt^{ISR}$.
%
%
The samples of 1-jet \zpj events, gained from the  PYTHIA simulation
of $5\cdot10^6$  signal \zpj events in two $\Ptz$
intervals 45 -- 55 and 100 -- 120 $GeV/c$, will be used here.
The observables defined in Section 2  will be
restricted here by Selection 1 cuts (\ref{eq:sc1}) -- (\ref{eq:sc6})
of Section 2.2 with $\Pt^{clust}_{CUT}=30 ~GeV/c$. $\Pt^{out}_{CUT}$ is not limited here.

\begin{figure}[htbp]
\vspace{-3.0cm}
  \hspace{.0cm} \includegraphics[width=16cm]{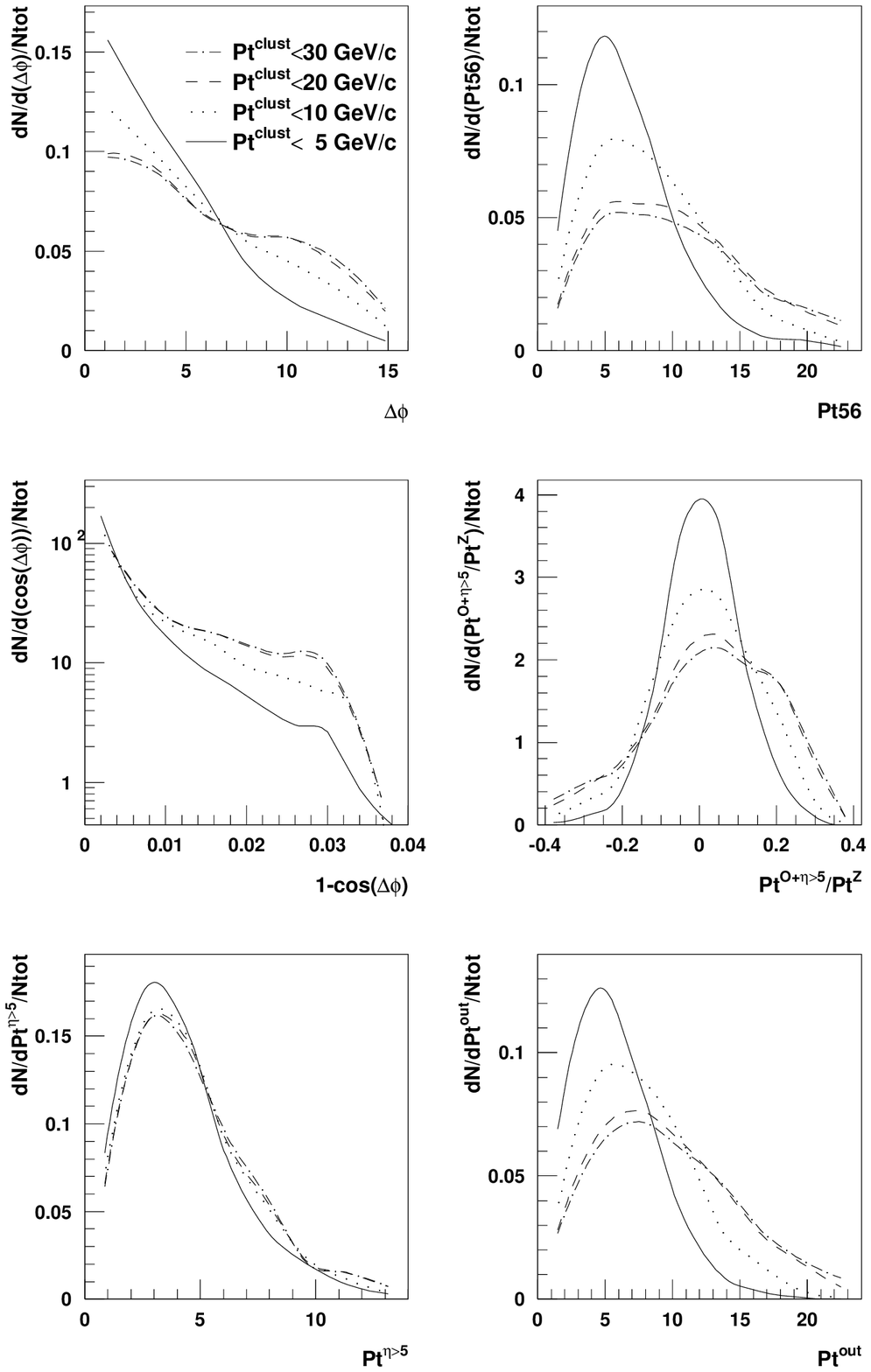}
  \vspace{-0.5cm}
    \caption{\hspace*{0.0cm} LUCELL algorithm, $\dphi\lt15^\circ$,
$45\lt\Pt^{Z}\lt55\, GeV/c$. Selection 1.}
 \label{fig:40b-luc}
\end{figure}

\begin{figure}[htbp]
 \vspace{-3.0cm}
  \hspace{.0cm} \includegraphics[width=16cm]{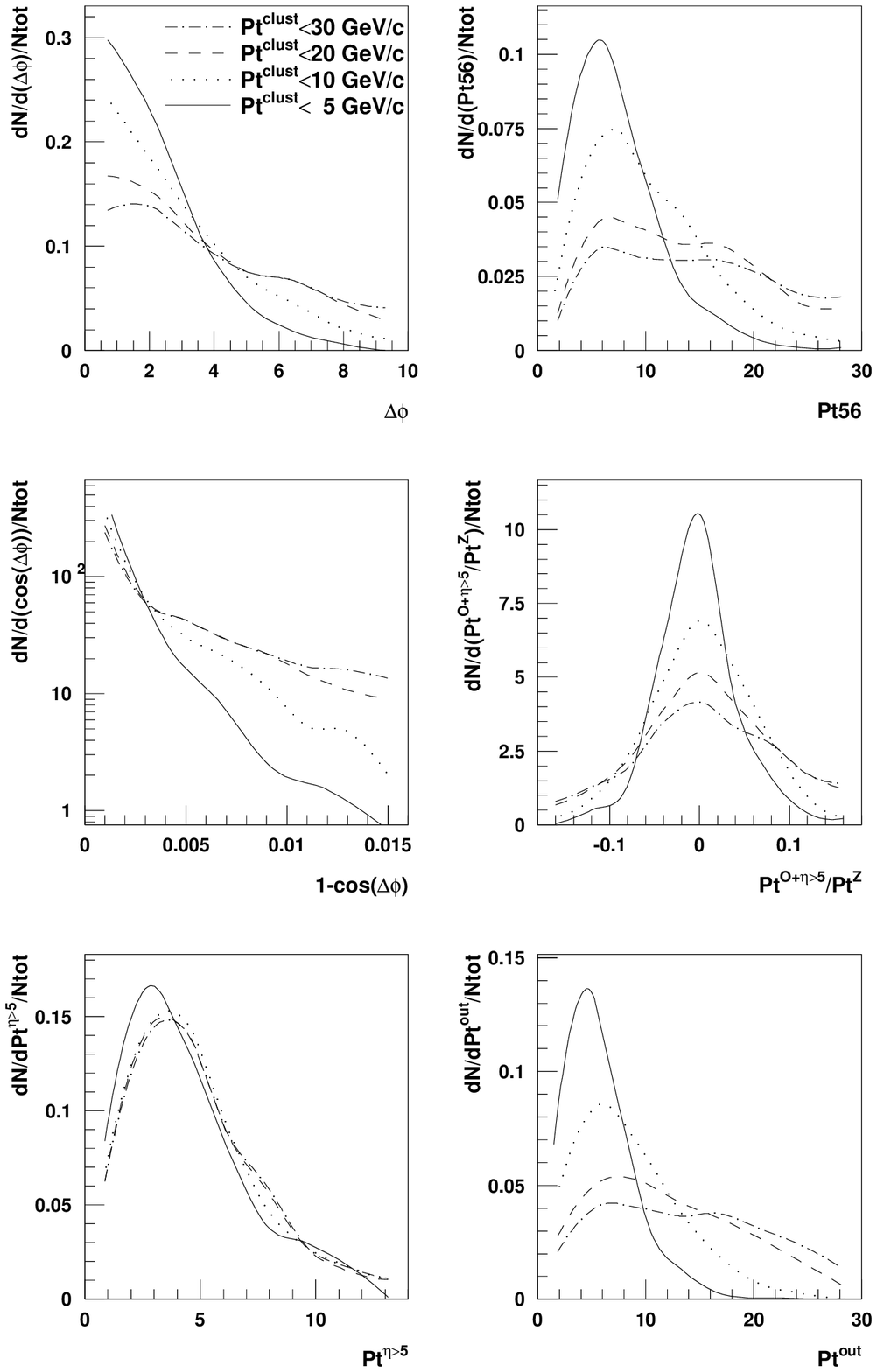}
  \vspace{-0.5cm}
    \caption{\hspace*{0.0cm} LUCELL algorithm, $\dphi\lt15^\circ$,
$100\lt\Pt^{Z}\lt120\, GeV/c$. Selection 1.}
    \label{fig:100b-luc}
\end{figure}

The influence of the $\Pt^{clust}_{CUT}$ variation 
on the distribution of some important physical variables 
is shown in Fig.~\ref{fig:40b-luc} for $45\lt\Ptz\lt55~ GeV/c$ and 
in Fig.~\ref{fig:100b-luc} for $100\lt\Ptz\lt120~ GeV/c$. Besides of distributions for three auxiliary 
variables $\Pt56$, $\Pt^{\eta\gt5}$, $\Pt^{out}$ (defined by (2), (16), (18))
we present distributions for $\Db$~ and $(1\!-\!cos\dphi)$ which
define the right-hand side of equation (\ref{eq:sc_bal}).
The distribution of the back-to-back $\dphi$ angle (\ref{eq:sc4}), defining the second
variable $(1\!-\!cos\dphi)$, is also presented in Figs.~\ref{fig:40b-luc}, ~\ref{fig:100b-luc}.

The $\Pt56$ variable and both components defining \ptzj disbalance,
$(1-cos\dphi)$ and $\Db$, as well as two others variables, 
$\Pt^{out}$ and  $\dphi$, show a tendency to become smaller (as the mean values as the widths 
of distributions) by restricting an upper limit on the $\Pt^{clust}$ value (see  
also tables of Appendices 2--5
). It means that the precision of jet energy setting may increase with decreasing
$\Pt^{clust}_{CUT}$. 
The origin of this improvement becomes clear from the
$\Pt{56}$ density plot which demonstrates ISR suppression
(or $\Pt^{ISR}$) as a more restrictive cut is imposed on $\Pt^{clust}$.

Comparison of Fig.~\ref{fig:40b-luc} (for $~45\!\lt\Ptz\!\lt55 ~GeV/c$) and Fig.~\ref{fig:100b-luc}
(for $~100\!\lt\Ptz\!\lt120 ~GeV/c$) shows that $\Delta\phi$ as a degree of
back-to-backness of $Z^0$ boson and jet $\Pt$ vectors in the $\phi$-plane
decreases with increasing $\Ptz$.  At the same time $\Pt^{ISR}$ distribution becomes wider, while
the $\Pt^{\eta\gt5}$ and $\Pt^{out}$ distributions practically do not depend on
$\Ptz$ (see for details Appendices 2--5).

It should be mentioned that the results presented in Figs.~\ref{fig:40b-luc}
and \ref{fig:100b-luc} were obtained with the LUCELL jetfinder of PYTHIA
\footnote{The results obtained with all jetfinders and
\ptzj ~balance will be discussed in Sections  7 in more detail.}.

\begin{center}
\begin{figure}[htbp]
 \vspace{-3.0cm}
  \hspace{.0cm} \includegraphics[width=16cm]{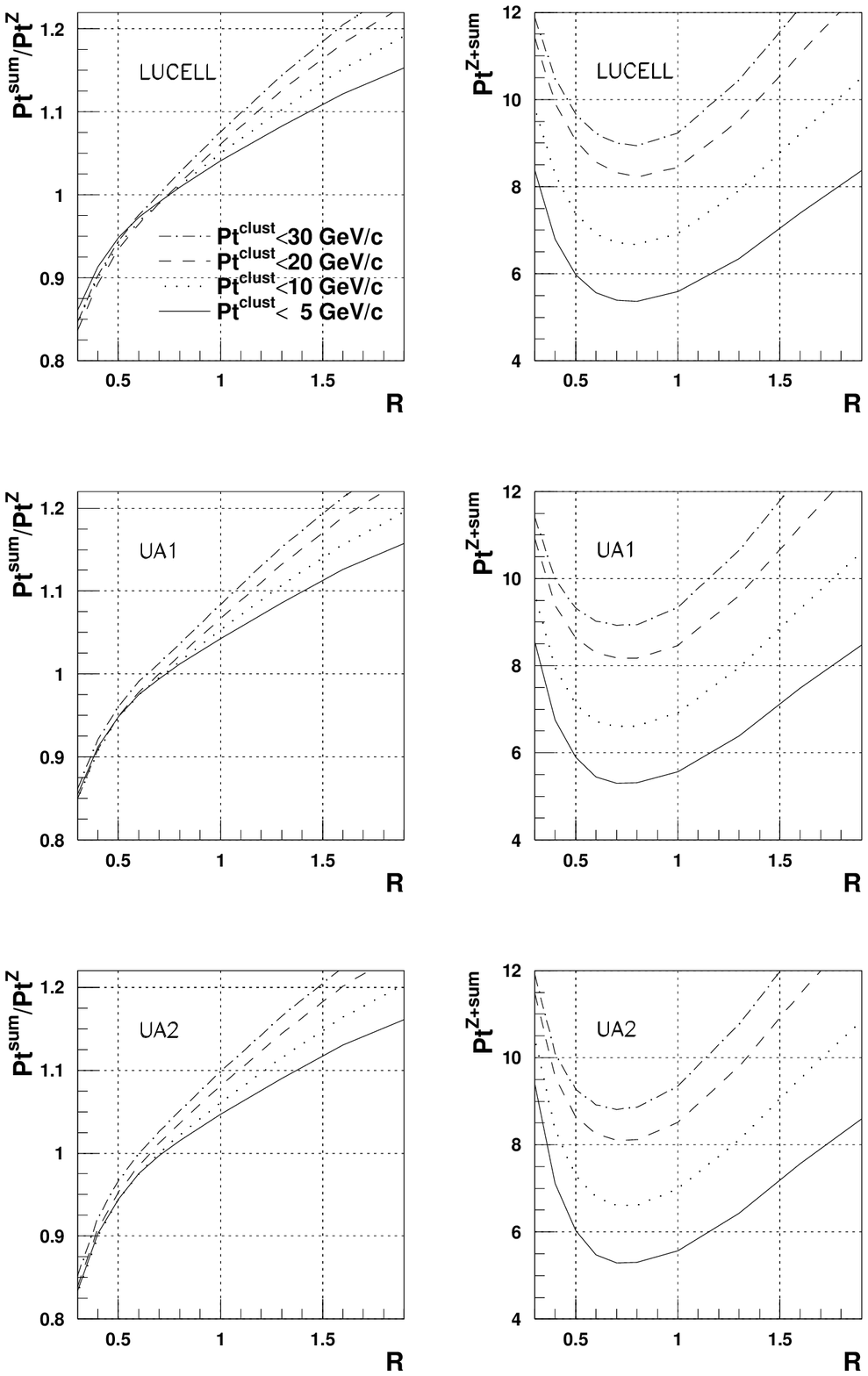}
  \vspace{-0.5cm}
    \caption{\hspace*{0.0cm} LUCELL, UA1 and UA2 algorithms, $\dphi\lt15^\circ$,
$45\lt\Pt^{Z}\lt55\, GeV/c$.}
    \label{fig:40b-ptr}
\end{figure}
\begin{figure}[htbp]
 \vspace{-3.0cm}
  \hspace{.0cm} \includegraphics[width=16cm]{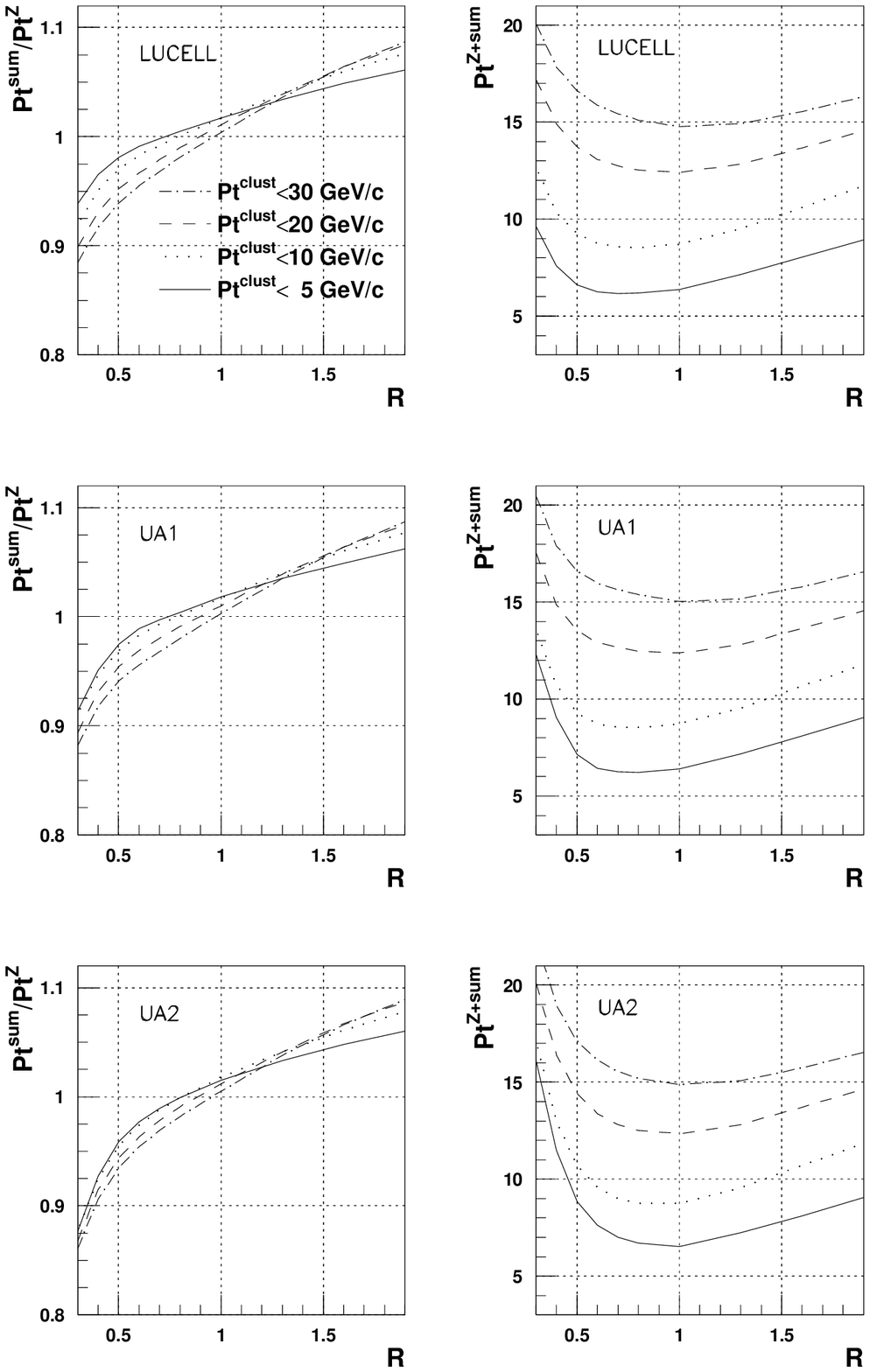}
  \vspace{-0.5cm}
    \caption{\hspace*{0.0cm} LUCELL, UA1 and UA2 algorithms, $\dphi\lt15^\circ$,
$100\lt\Pt^{Z}\lt120\, GeV/c$.}
    \label{fig:100b-ptr}
\end{figure}
\end{center}

~\\[-20mm]

\noindent


\subsection{$\Pt$ distribution inside and outside of a jet.}

Now let us see how the space outside the jet
may be populated by $\Pt$ in the \zpj HB events. For this
purpose we calculate a vector sum $\vec{\Pt}^{sum}$ of individual transverse
momenta of 
the calorimeter cells included by a jetfinder into a jet and of cells in a larger
volume that surrounds a jet. In the latter case this procedure can 
be viewed as straightforward enlarging of the jet radius in the $\eta -\phi$ space.

The plots that present the ratio  $\Pt^{sum}/{\Ptz}$
as a function of the distance $R(\eta,\phi)$ counted from a jet
gravity center towards its boundary and further into the space outside a jet
are shown in the left-hand columns of Figs.~\ref{fig:40b-ptr} and \ref{fig:100b-ptr} for
two $\Ptz$ intervals ($45\lt\Ptz\lt55 ~GeV/c$ and $100\lt\Ptz\lt120 ~GeV/c$)
and three jetfinding algorithms (UA1, UA2 and LUCELL).

From these figures we see that the space surrounding the jet is
in general far from being empty. We also see that the average value of
$\Pt^{sum}$ increases with increasing volume around a jet
and it exceeds $\Ptz$ at $R=0.7-0.8$ (see Figs.~\ref{fig:40b-ptr} and \ref{fig:100b-ptr}).

From the right-hand columns of Figs.~11 and 12 we see that 
the vector disbalance measure 
~\\[-5pt]
\begin{equation}
\Pt^{Z+sum}=
\left|\vec{\Pt}^Z+\vec{\Pt}^{sum}\right|
\end{equation}
achieves its minimum again at $R\approx 0.7-0.8$ for all jetfinding algorithms.
(The minimum of the vector sum $\Pt^{Z+sum}$ can serve as an illustration of 
the $\Ptz\!-\!\Pt^{jet}$ disbalance minimum.)                        

The value of $\Pt^{Z+sum}$ (as well as $\Pt^{sum} /\Ptz$) continues to grow rapidly for 
$40\lt\Ptz\lt50~ GeV/c$ and more slowly for $100\lt\Ptz\lt120 ~GeV/c$ with increasing $R$ 
after the point $R=0.7-0.8$ 
(see Figs.~\ref{fig:40b-ptr} and \ref{fig:100b-ptr}). This means that at higher $\Ptz$
(or $\Pt^{Jet}$) the topology of \zpj events becomes more
distinct and we get a clearer picture of an ``isolated" jet. This feature
clarifies the motivation of introducing the ``Selection 2'' criteria
in Section 2.2 for selection of events with isolated jets.

\section{Dependence of the  disbalance between \ptzj
on the $\Pt^{clust}_{CUT}$ and $\Pt^{out}_{CUT}$ parameters.}

Here we shall study in detail a dependence of the $\Ptz-\Pt^{Jet}$ disbalance
on the values of $\Pt^{clust}_{CUT}$ and
$\Pt^{out}_{CUT}$. For this aim the four samples of \zpj events
described in the beginning of Section 4 were used. 
%
%

The mean values of the most important variables used in our analyses  
that reflect the main features of \zpj events with the jet completely contained in 
the Barrel region, i.e. ``HB events'' (see Section 5) are given in the tables of Appendices 2--5.

Appendix 2 contains the tables
for events inside $45\lt\Ptz\lt 55~GeV/c$ interval. In these tables we present the values of
interest found with UA1, UA2 and LUCELL jetfinders for three different Selections
mentioned in Section 2.2. Each page corresponds to a definite value of $\Delta \phi$ 
as a measure of deviation from the absolute
back-to-back orientation of $\vec{\Pt}^Z$ and $\vec{\Pt}^{Jet}$ vectors.
The first four pages of each Appendix contain the information
about variables that characterize the $\Ptz$ -- $\Pt^{Jet}$ balance
for events passed the cuts (\ref{eq:sc1})--(\ref{eq:sc6}) (Selection 1).

On the fifth page of each of Appendices 2--5 we present Tables
13 -- 15 (for the cut $\dphi\lt15^{\circ}$)
that correspond to Selection 2 (see Section 2.2).
We have limited $\epsilon^{jet}\leq 9\%$ for
$45\lt\Ptz\lt 55$ with a gradual change to $\epsilon^{jet}\leq 3\%$ for $\Ptz\geq 100 ~GeV/c$.
The best result for UA2 in the case of $45\lt\Ptz\lt 55$ is obtained with $\epsilon^{jet}\leq 6\%$
instead of the cut $\epsilon^{jet}\leq 9\%$ chosen for UA1 and LUCELL algorithms
\footnote{In \cite{BKS_P1} -- \cite{BKS_P5} the Selection 2 criterion was considered 
with a more severe cut $\epsilon^{jet} \leq 2\%$.}.
The results obtained with Selection 3
are given on the sixth page of Appendices 2--5
\footnote{Selection 3 (see Section 2.2) leaves
only those events in which jets are found simultaneously by UA1, UA2 and LUCELL jetfinders
i.e. events with jets having up to a good accuracy equal coordinates of
the center of gravity, $\Pt^{jet}$ and $\phizj$.}.
%
while on the seventh page Selection 3 is used to find jets found 
simultaneously by UA1 and LUCELL jetfinders only.

The columns in tables of Appendices 2--5 correspond to five 
values of $\Pt^{clust}_{CUT}=30,\ 20,\ 15,$ $10$ and $5 ~GeV/c$.
The upper lines of these tables contain the expected numbers 
$N_{event}$ of ``HB events''
for the integrated luminosity $L_{int}=10\;fb^{-1}$. 

In the next four lines of the tables we put the values of $\Pt56$,
$\Delta \phi$, $\Pt^{out}$, $\Pt^{\eta \gt 5}$
defined by formulas (2), (9), (18) and (16), respectively, and
averaged over the events selected with a chosen $\Pt^{clust}_{CUT}$ value.
From the tables we see, firstly, that the averaged values of
$\Pt^{\eta\gt5}$ show very weak dependence on it (practically constant)
\footnote{Compare with Figs.~7 and 8.},
what is in complete agreement with behavior of these variables 
in the case of \gpj events \cite{BKS_P3}. At the same time, 
the values of $\Pt56$, $\Delta \phi$, $\Pt^{out}$ decrease fast
with decreasing $\Pt^{clust}_{CUT}$.
The $\Pt56$ variable (non-observable one) that serves, according to (2), as
measure of the initial state radiation transverse momentum $\Pt^{ISR}$,
i.e. one of the main source of the $\Pt$ disbalance in the 
subprocesses (2a) and (2b).
So, variation of $\Pt^{clust}_{CUT}$ from $30$ to
$10 ~GeV/c$ for $\dphi\lt15^\circ$
leads to suppression of the $\Pt56$ value
(or $\Pt^{ISR}$) approximately by $\approx 40-45\%$ for all $\Ptz$. 

The following three lines (from 6-th to 8-th) show the average values of the variables
$\Zpart$,
$\Jpart$,
$\ZJ$ (here J$\equiv$Jet).
These lines correspond to the relative
$\Pt$ balance at the $Z^0$--parton level (final state of the fundamental subprocess $2\to 2$), 
the relative difference of the parton $\Pt$ and the jet $\Pt$ (parton hadronization
effect) and the relative $\Pt$ balance of the jet and $Z^0$ boson.

The lines 9 and 10 include the averaged values of $\Db/\Ptz$ and $\,(1-cos(\dphi))$
that appear on the right-hand side of the $\Pt$-balance equation (\ref{eq:sc_bal}).

As a rule, the value of $\left<1\!-\!cos(\dphi)\right>$ is smaller than the value of
$\left<\Db/\Ptz\right>$ for the cut $\dphi\lt15^\circ$ and tends to decrease more with
growing energy. So, we can conclude that the main source of the $\Pt$
disbalance in the \zpj system is defined by the term $\Db/\Ptz$.

The~ following~ line~ contains~ the averaged~ values~ of~ the standard~ deviations~ of
$\ZJ (\equiv Db[Z,J])$. The values of  this variable
drop approximately by a factor of two (and even more for all intervals with $\Ptz\gt100 ~GeV/c$) 
while moving from $\Pt^{clust}_{CUT}=30 ~GeV/c$ to $\Pt^{clust}_{CUT}=5 ~GeV/c$
for all jetfinding algorithms.

The last lines of the tables present the number of generated events, i.e. entries left after cuts.

A decrease in $\Pt^{clust}_{CUT}$ leads to
a decrease in the $(\Ptz-\Pt^J)/\Ptz$
ratio (mean values as well as standard deviations), i.e. we select the events that can be used to
improve the jet energy calibration accuracy. For instance, in the
case of $70\lt\Ptz\lt85 ~GeV/c$ the mean value of
$(\Ptz\!-\!\Pt^J)/\Ptz$ drops from $4.5\!-\!4.9\%$ to
$0.9\!-\!1.5\%$ (see Tables 4, 6 of Appendix 3) and in the
case of $100\lt\Ptz\lt120 ~GeV/c$ the mean value of this variable
drops from $3.4-3.7\%$ to less than $0.6-1.0\%$ for UA1 and LUCELL jetfinders
(Tables 4, 6 of Appendix 4).
A worse situation is seen for the $45\lt\Ptz\lt 55 ~GeV/c$ interval, where
the disbalance changes, e.g. for LUCELL algorithm, as $2.1\to 1.8\%$.
Meantime, RMS values with the same variations of $\Pt^{clust}_{CUT}$ (from 30 to 10 $GeV/c$)
decrease by $40-50\%$.

After imposing the jet isolation requirement (see Tables 13 -- 15 of Appendices 2--5)
we observe that for $\Ptz\geq100 ~GeV/c$ the mean values of
$(\Ptz\!-\!\Pt^J)/\Ptz$ are contained inside the $1\%$ window
for any $\Pt^{clust}$. For $45\lt\Ptz\lt55 ~GeV/c$
we see that with Selection 1 $\Pt^{clust}_{CUT}$ works more effectively than in the case of Selection 1.
%
%
Thus, $\Pt^{clust}_{CUT}=15 ~ GeV/c$
allows to reduce $(\Ptz\!-\!\Pt^J)/\Ptz$ to less than $1\%$ level
for all algorithms. The Selection 2 criterion 
leaves quite a sufficient number of events with a jet contained completely in the barrel region:
about 10~000 -- 18~000
\footnote{the lower value corresponds to UA2 algorithm for which the stricter jet isolation cut
was used}
for $45\lt\Ptz\lt 55 ~GeV/c$ with $\Pt^{clust}_{CUT}=15 ~GeV/c$  
and about 4~500  for $100\lt\Ptz\lt120 ~GeV/c$  with $\Pt^{clust}_{CUT}=20 ~GeV/c$ 
at $L_{int}=10 ~fb^{-1}$ (see Tables 13 -- 15 of Appendices 2, 3).

The analogous results for Selection 3 are presented in Tables 16--18 of Appendices 2--5.
Let us consider first the most difficult interval $45\lt\Ptz\lt55 ~GeV/c$.
From the tables of Appendix 2 one can see that this selection leads to approximately $20\%$ reduction
of the number of selected events as compared with the case of Selection 2.
A combined usage of all three jetfinders (Tables 16--18) worsen the balance values.
A requirement of simultaneous jet finding by only UA1 and LUCELL algorithms
practically does not change values of the $\Ptz-\Pt^{Jet}$
 balance and other variables, presented in Tables 19, 20,
as compared with the case of Selection 2 and gives a better result
(from point of view of the $\Ptz-\Pt^{Jet}$ balance values as well as from point of view of
the number of selected events) as compared with the case of combined usage of all three 
jetfinders for this aim. 
This fact stresses a good compatibility of UA1 and LUCELL jetfinders.
%
%
For other considered $\Ptz$ intervals  UA1, UA2 and LUCELL algorithms give more or 
less close results and a passage to Selection 3 does not worsen a situation.

We also can note that Selections 2 and 3, besides improving the \ptzj balance value,
are important for selecting events with a clean jet topology and rising the confidence level 
of a jet determination.

The influence of a wide variation of cuts $\Pt^{clust}_{CUT}$ and  $\Pt^{out}_{CUT}$ on \\
\noindent
(a) the number of selected events (for $L_{int}=10\,fb^{-1}$),\\
(b) the mean value of $F\equiv(\Pt^{Z}\!-\!\Pt^{Jet})/\Pt^{Z}$ and \\
(c) the standard deviation value $\sigma (F)$\\
is presented in rows and columns of Tables 1--9 for Selection 1 of Appendix 6.
The set of selection cuts (4)--(10) (Section 2.2) was applied to preselect
\zpj events 
for the tables of  Appendix 6.
The jets (as well as clusters) in these events, unlike the the jets in the events analyzed 
in Appendices 2--5, {\it were found by LUCELL jetfinder  
for the whole $\eta$ region $|\eta^{jet}|<5.0$}.

Tables 1--3 of Appendix 6 correspond to the \zpj events selection in the interval 
$40\leq\Ptz\leq70~GeV/c$ 
Tables 4--6 to that for $70\leq\Ptz\leq100~GeV/c$  and
Tables 7--9 to that for $100\leq\Ptz\leq140~GeV/c$.


We see that the restriction of $\Pt^{clust}$ and $\Pt^{out}$ are necessary to 
improve the jet energy setting accuracy. So, Tables 2 (for $40\leq\Ptz\leq70~GeV/c$) and 
8 (for $100\leq\Ptz\leq140~GeV/c$) of Appendix 6 show that the mean values of the fraction
$F\equiv \Fptzj$ decreases with variation of the two cuts 
from $\Pt^{clust}_{CUT}=30~ GeV/c$ and $\Pt^{out}_{CUT}=1000 ~GeV/c$ (i.e. without limits)
to $\Pt^{clust}_{CUT}=10~ GeV/c$ and $\Pt^{out}_{CUT}=10 ~GeV/c$
as 0.049  to 0.018 and as 0.036 to 0.012, respectively.
At the same time this restriction noticeably decreases 
the width of the Gaussian $\sigma (F)$ (see Tables 3, 6 and 9 of Appendix 6). 
So, it drops from 0.200 to 0.103 for $40\leq\Ptz\leq70~GeV/c$  and
 from 0.138 to 0.066 for $100\leq\Ptz\leq140~GeV/c$ (i.e. about in a factor of two)
for the same variation of $\Pt^{clust}_{CUT}$ and $\Pt^{out}_{CUT}$.

Again, the reason is caused by the term $\Db/\Ptz$ of the $\Pt$-balance equation (19)
(as we noted above, the contribution of ($1-cos\dphi$) to the \PTzj disbalance is
negligibly small).  
This term can be decreased by decreasing
$\Pt$ activity in the space {\it out of} the \zpj system, i.e. by limiting 
$\Pt^{clust}$ and $\Pt^{out}$.

The numbers of events at the integrated luminosity $L_{int}= 10~fb^{-1}$
for different $\Pt^{clust}_{CUT}$ and $\Pt^{out}_{CUT}$
are given in Tables 1, 5 and 9 of Appendix 6. 
One can see that even with such strict $\Pt^{clust}_{CUT}$ and $\Pt^{out}_{CUT}$ values as $10 ~GeV/c$ 
for both, for example, we would have 
69~600,  18~100 and 6~860 for $40\leq\Ptz\leq70~GeV/c$, $70\leq\Ptz\leq100~GeV/c$ and
$100\leq\Ptz\leq140~GeV/c$ respectively.

In addition, we present in Tables 10--18 of Appendix 6 the results obtained with Selection 2.
They contain the information analogous to that in Tables 1--12
but for the case of imposing jet isolation requirement: $\epsilon^{jet}=8\%$ at
$40\leq\Ptz\leq70~GeV/c$ and $\epsilon^{jet}=5\%$ at  $70\leq\Ptz\leq100~GeV/c$
and $100\leq\Ptz\leq140~GeV/c$.
%
%
From these tables we see that with the same (and with even weaker) cuts
$\Pt^{clust}_{CUT}=\Pt^{out}_{CUT}=10 ~GeV/c$ one can obtain a 
much better fractional balance $F$, less than $1\%$ for all $\Ptz$ intervals
(with almost the same values of $\sigma(F)$), at the statistics of
about 50~800, 13~200 and 6~150 events for intervals 
$40\leq\Ptz\leq70~GeV/c$,  $70\leq\Ptz\leq100~GeV/c$ and $100\leq\Ptz\leq140~GeV/c$,  
respectively

\begin{figure}[htbp]
\vskip-1mm
  \hspace*{10mm} \includegraphics[width=14cm,height=15.9cm]{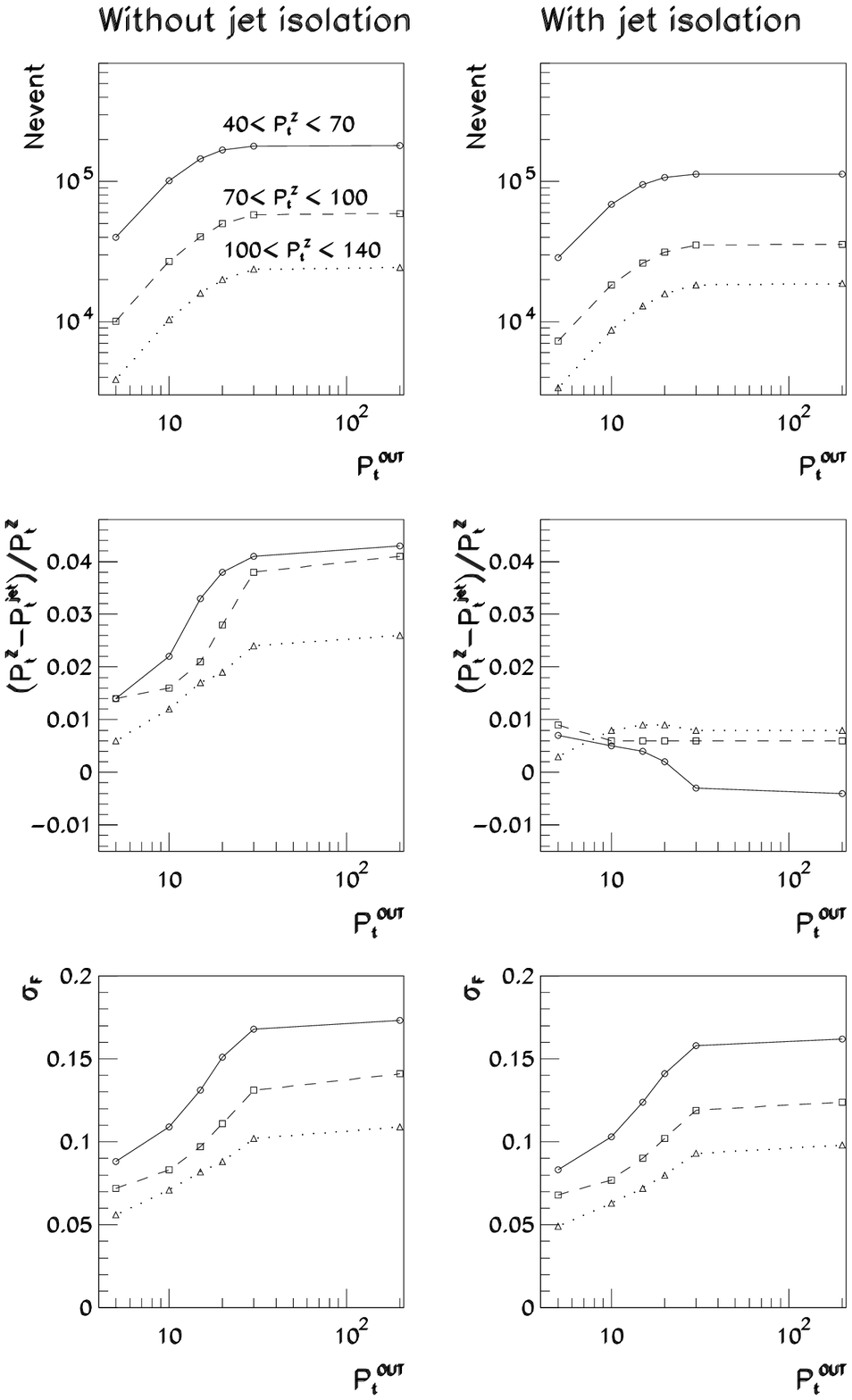} 
\vskip-5mm
    \caption{\hspace*{0.0cm} Number of events at $L_{int}=10 ~fb^{-1}$, mean value
$\Fptzj (\equiv F)$, its standard deviation $\sigma_F$ as a function of $\Pt^{out}_{CUT}$ value.
$\Pt^{clust}_{CUT}$ value is limited by $20~GeV/c$. Full line corresponds to the event selection
with $40\leq\Ptz\leq70~GeV/c$, dashed line to that with $70\leq\Ptz\leq100~GeV/c$ and
dotted line to that with $100\leq\Ptz\leq140~GeV/c$ ($\epsilon^{jet}=8\%, 5\%, 5\%$ 
in these $\Ptz$ intervals, respectively).}
\label{fig:mu-sig}
\end{figure}

The behavior of number of the selected events for $L_{int}=10\,fb^{-1}$,
the mean values of $\Fptzj$ and its standard deviation $\sigma (F)$ as a function
of $\Pt^{out}_{CUT}$ for $\Pt^{clust}_{CUT}=20~GeV/c$
is displayed in Fig.~\ref{fig:mu-sig} for non-isolated  (left-hand column) and
isolated jets with $\epsilon^{jet}=8\%$ at
$40\leq\Ptz\leq70~GeV/c$ and $\epsilon^{jet}=5\%$ at  $70\leq\Ptz\leq100~GeV/c$
and $100\leq\Ptz\leq140~GeV/c$ (right-hand column).

\newpage

\section{The study of background suppression.}                   

In principle, there is a probability, that some combination of muons in
the events, based on the QCD subprocesses with much larger cross sections 
(by about 5 orders of magnitude) than ones of  
the signal subprocesses,
can be registered as the $Z^0$ signal. This type of background we call as ``combinatorial background''.
To study a rejection possibility of such type of events by about 40 million events
with a mixture of all QCD and SM subprocesses with large
cross sections existing in PYTHIA 
\footnote{(namely, ISUB=11--20, 28--31, 53, 68)} 
including also the signal subprocesses (2a) and (2b)
were generated. Three generations were performed with
different minimal $\Pt$ of the hard $2\to 2$ subprocess
\footnote{i.e $CKIN(3)$ parameter in PYTHIA}
$\pth$ values: $\pth$= 40, 70 and 100 $GeV/c$. 
The cross sections of different subprocesses 
serve in simulation as weight factors and, thus, determine
the final statistics of the corresponding physical events.
The generated events were analyzed by use of the cuts given in Table \ref{tab:sb_mc0}
(see also Section 2.2).
\\[-2mm]
\begin{table}[h]
\caption{List of the applied cuts used in Tables \ref{tab:sb_mc1}, \ref{tab:sb_mc2}.}
\begin{tabular}{lc} \hline
\label{tab:sb_mc0}
~~ \hspace*{148mm} ~~\\[-3mm]
\hspace*{14mm} {\bf 0}. Total number of $\mu^+ \mu^-$-- pairs (No selection); ~~~~~~~~~~~~~~~~~~~~~~~~~~~~~~~~~~~\\[1pt]
\hspace*{15mm}{\bf 1}. $\Pt^{\mu}>10~~GeV/c$, $|\eta^{\mu}|<2.4$; \\[1pt]   
\hspace*{15mm}{\bf 2}. $|M_{\bf Z}-M_{inv}^{ll}|<20~~GeV/c^2$; \\[1pt]
\hspace*{15mm}{\bf 3}. 1 jet events selected; \\[1pt]
\hspace*{15mm}{\bf 4}. $\Pt^{isol}/\Pt^\mu \leq 0.10,~\Pt^{ch}<2 ~~GeV/c $;\\[1pt]
\hspace*{15mm}{\bf 5}. $|M_{\bf Z}-M_{inv}^{ll}|<5~~GeV/c^2$; \\[1pt]
\hspace*{15mm}{\bf 6}. $\dphi<15^\circ$. \\\hline 
\end{tabular}
\end{table}

To trace the effect of their application let us consider first the case of one
(intermediate) energy, i.e. the generation with $\pth$=70 $GeV/c$.
Each line of Table \ref{tab:sb_mc1} corresponds to the respective cut of Table \ref{tab:sb_mc0}.
The numbers in columns ``Signal'' and ``Bkgd'' show the number of signal and (combinatorial)
background events remained after a cut. Column ``$Eff_{S(B)}$'' demonstrates the efficiency 
of a cut.
The efficiencies $Eff_{S(B)}$ (with their errors) are defined as a ratio
of the number of signal (background) events that passed under a cut
(1--6) to the number of the preselected events after the furst cut of Table \ref{tab:sb_mc0}
(The number of events after the first cuts is taken as $100\%$).
\begin{table}[h]
\begin{center}
\vskip-1mm
\caption{A demonstration of cut-by-cut efficiencies and $S/B$ ratios for generation with
$\pth$=70 $GeV/c$.}
\small
\vskip0.1cm
\label{tab:sb_mc1}
\begin{tabular}{||c||c|c|c|c|c||}                  \hline \hline
Selection & Signal & Bkgd & $Eff_S(\%)$ &$Eff_B(\%)$ & $S/B$ \\\hline \hline
\rowcolor[gray]{\coltab}%
 0  &  401 & 850821 &           &             & $5\cdot10^{-4}$ \\\hline \hline  
 1  &  245 &  15842 &100.00$\pm$0.00 &100.00$\pm$0.000 & 0.02 \\\hline \hline   
 2  &  226 &    467 & 92.24$\pm$8.51 &  2.948$\pm$0.138 & 0.5 \\\hline 
 3  &   99 &     12 & 40.41$\pm$4.81 &  0.076$\pm$0.022 & 8.3 \\\hline 
 4  &   81 &     10 & 33.00$\pm$4.24 &  0.063$\pm$0.020 & 8.1 \\\hline 
\rowcolor[gray]{\coltab}%
 5  &   72 &      4 & 29.39$\pm$3.94 &  0.025$\pm$0.013 &18.0 \\\hline 
\rowcolor[gray]{\coltab}%
 6  &   62 &      0 & 25.31$\pm$3.60 &  0.000$\pm$0.000 & --  \\\hline \hline        
\end{tabular}
\vskip5mm
\caption{Values of efficiencies and $S/B$ ratios for generations with
$\pth$=40, 70 and 100 $GeV/c$.}
\small
 \vskip0.1cm
\label{tab:sb_mc2}
\begin{tabular}{||c||c|c|c|c|c|>{\columncolor[gray]{\coltab}}c||}                  \hline \hline
$\pth$& Cuts& Signal& Bkgd &$Eff_S(\%)$&$Eff_B(\%)$&$S/B$\\\hline \hline
40 &Preselection (1)  & 89&  1090&100.00$\pm$0.00 &100.00$\pm$0.00 &0.08\\\cline{2-7}
$(GeV/c)$ & Main ($1-5$) & 30&     0& 33.71$\pm$7.12 & 0.00$\pm$0.00  & -- \\\hline \hline
70 &Preselection (1)  &245& 15842&100.00$\pm$0.00 &100.00$\pm$0.00  &0.02\\\cline{2-7}
$(GeV/c)$ & Main ($1-5$) & 72&     4& 29.39$\pm$3.94 & 0.025$\pm$0.013 &18.0\\\hline \hline
100 &Preselection (1) &497& 37118&100.00$\pm$0.00 &100.00$\pm$0.00  &0.01\\\cline{2-7}
$(GeV/c)$ & Main ($1-5$) &127&     4& 25.55$\pm$2.54 & 0.011$\pm$0.005 &31.8\\\hline \hline
\end{tabular}
\end{center}
\end{table}

We see from Table \ref{tab:sb_mc1} that initial ratio of $\mu^+\mu^-$ pairs in
signal and background events is very small ($5\cdot10^{-4}$)
\footnote{That is mainly due to the huge difference in the cross sections of \zpj events (from subprocesses
(2a), (2b)) and the QCD events.}.
%
A weak restriction of the muon transverse momentum and pseudorapidity
in the $1st$ selection increase $S/B$ by about 2 order 
(as $5\cdot10^{-4}\to2\cdot10^{-2}$).
The invariant mass criterion and one-jet events selection make $S/B=18.0$ and
the last criterion on the azimuthal angle between $Z^0$ and jet 
($\dphi<15^\circ$) suppresses the background events completely.

The information on other intervals (i.e. on the event generations with $\pth=40$ and $\pth=100~GeV/c$)
is presented in Table \ref{tab:sb_mc2}.
Line ``Preselection (1)'' corresponds to the first cuts in Table \ref{tab:sb_mc0} 
($\Pt^{\mu}>10~~GeV/c$, $|\eta^{\mu}|<2.4$) while line ``Main ($1-5$)'' corresponds to the result of
application of criteria from 1 to 5 of Table \ref{tab:sb_mc0}. 
{\it After application of all six criteria of Table \ref{tab:sb_mc0} 
we have observed no background events in all of the $\Ptz$ intervals} 
with the signal events selection efficiency of $25-33\%$.

The practical absence of a background to the \zpj events allow
to use them for an extraction of the gluon distribution in a proton $f^g_p(x,Q^2)$.

%
\section{Estimation of rates 
for gluon distribution determination  at the LHC using \zpj events.}

Many theoretical predictions for production of new particles
(Higgs, SUSY) at the LHC are based on model estimations of the gluon density behavior at
low $x$ and high $Q^2$. Thus, determining the proton gluon density $f^g_p(x,Q^2)$
for this kinematic region directly in LHC experiments would be obviously very useful. 

One of the channels for this determination is a high $\Pt$ direct photon production 
$pp\rightarrow \gamma^{dir} + X$ (see \cite{Au1}).
The region of high $\Pt$, reached by UA1 \cite{UA1}, UA2 \cite{UA2}, CDF \cite{CDF1} and
D0 \cite{D0_1} extends up to $\Pt \approx 60~ GeV/c$ and recently up to $\Pt= 105~ GeV/c$ \cite{D0_2}. 
These data together with the later ones (see references in \cite{Fer}--\cite{Fr1}) and recent
E706 \cite{E706} and UA6 \cite{UA6} results give
an opportunity for tuning the form of gluon distribution (see \cite{Au2}, \cite{Vo1}).
The rates and estimated cross sections of inclusive direct photon production at the LHC are given in 
\cite{Au1}.

A more promising process that can be used for measuring $f^g_p(x,Q^2)$ is 
$pp\rightarrow \gamma^{dir}\, +\, 1\,jet\, + \,X$
defined at the leading order by two QCD subprocesses 
$qg\to q+\gamma$ and $q\bar{q}\to g+\gamma$ was considered in \cite{BKS_GLU,GLU_BKGD}
(see also \cite{MD1} and for experimental results see \cite{ISR}, \cite{CDF2}).

Here to estimate a possibility of extraction of information on the gluon density in a proton
we shall consider the \zpj production process (\ref{eq:zpj})
(analogous to the \gpj process above),
where $Z^0$ boson decays to the muon pair, a signal from which can be perfectly
measured in the detector.

In the case of $pp\to Z^0/\gamma^{dir}+1~ jet+X$ for $\Pt^{jet}\geq30~GeV/c$ 
(i.e. in the region where $k_T$ smearing
effects are not important, see \cite{Hu2}) the cross section is
expressed directly in terms of parton distribution functions $f^a(x_a,Q^2)$
(see, for example, \cite{Owe}): \\[-15pt]
\begin{eqnarray}
\frac{d\sigma}{d\eta_1d\eta_2d\Pt^2} = \sum\limits_{a,b}\,x_a\,f^a(x_a,Q^2)\,
x_b\,f^b(x_b,Q^2)\frac{d\sigma}{d\hat{t}}(a\,b\rightarrow 1\,2)
\label{eq:cross_gl}
\end{eqnarray}
\noindent
where $x_{a,b}$ are defined by \\[-22pt]
\begin{eqnarray}
x_{a,b} \,=\,\Pt/\sqrt{s}\cdot \,(exp(\pm \eta_{1})\,+\,exp(\pm \eta_{2})).
\label{eq:x_gl}
\end{eqnarray}
We also used the following designations above:
$\eta_1=\eta^Z$, $\eta_2=\eta^{jet}$; ~$\Pt=\Ptz$;~ $a,b = q, \bar{q},g$; 
$1,2 = q,\bar{q},g,Z^0$.
Formula (\ref{eq:cross_gl}) and the knowledge of the results of independent measurements of
$q, \,\bar{q}$ distributions \cite{MD1} allow the gluon  distribution $f^g_p(x,Q^2)$
to be determined with an account of the selection efficiencies of \zpj events.

In Table~\ref{tab:B30} we present the $Q^2 (\equiv(\Ptz)^2)$ and $x$
distribution (with $x$ defined by (\ref{eq:x_gl})) of the number of all events,
i.e. the events, based on the subprocesses $qg\to Z^0+q$ and $q\bar{q}\to g+Z^0$ 
(with the decay $Z^0\to\mu^+\mu^-$) for integrated luminosity $L_{int}=20 ~fb^{-1}$.
These events satisfy the cuts (4)--(12) of Section 2.2 with the parameter values:\\[-17pt]
\begin{eqnarray}
|\eta^{jet}|<5.0,\quad \Pt^\mu_{max} \geq 20~GeV/c, \quad \dphi<15^{\circ}, 
\quad \Pt^{clust}_{CUT}=10~GeV/c,\quad  \Pt^{out}_{CUT}=10~GeV/c.
\label{eq:par_gl}
~\\[-17pt]
\nonumber
\end{eqnarray}

\begin{figure}[htbp]
\vskip-17mm
   \hspace{10mm} \includegraphics[width=.90\linewidth,height=79mm,angle=0]{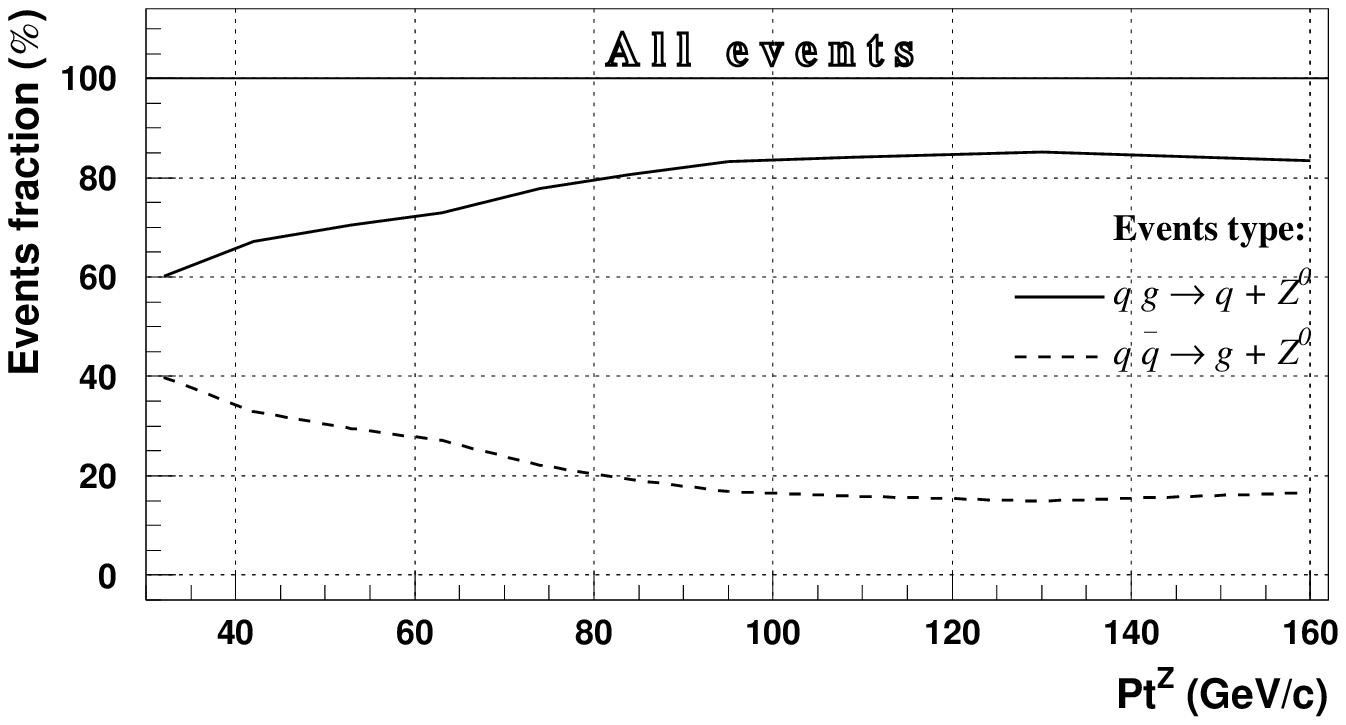}
\vskip-9mm
\caption{The contributions of the events originated from the subprocesses (2a) and (2b) 
as a function of $\Ptz$. Full line corresponds to the ``$qg\to q+Z^0$'' events, 
dashed line -- to the ``$q\bar{q}\to g+Z^0$'' events.}
\label{fig:procs}
\end{figure}

The contributions (in $\%$) of the events originated from the subprocesses (2a) and (2b) 
(and passed the cuts (4)--(12) of Section 2.2) 
as functions of $\Ptz$ are presented in Fig.~\ref{fig:procs}. 
From this figure one can see that the contribution of the events from the Compton
scattering (2a) varies from
$67\%$ at $\Ptz\approx 40~GeV/c$ to $85\%$ at $\Ptz\approx 120~GeV/c$.

\begin{figure}[h]
   \vskip-2mm
   \hspace{37mm} \includegraphics[width=.53\linewidth,height=7.9cm,angle=0]{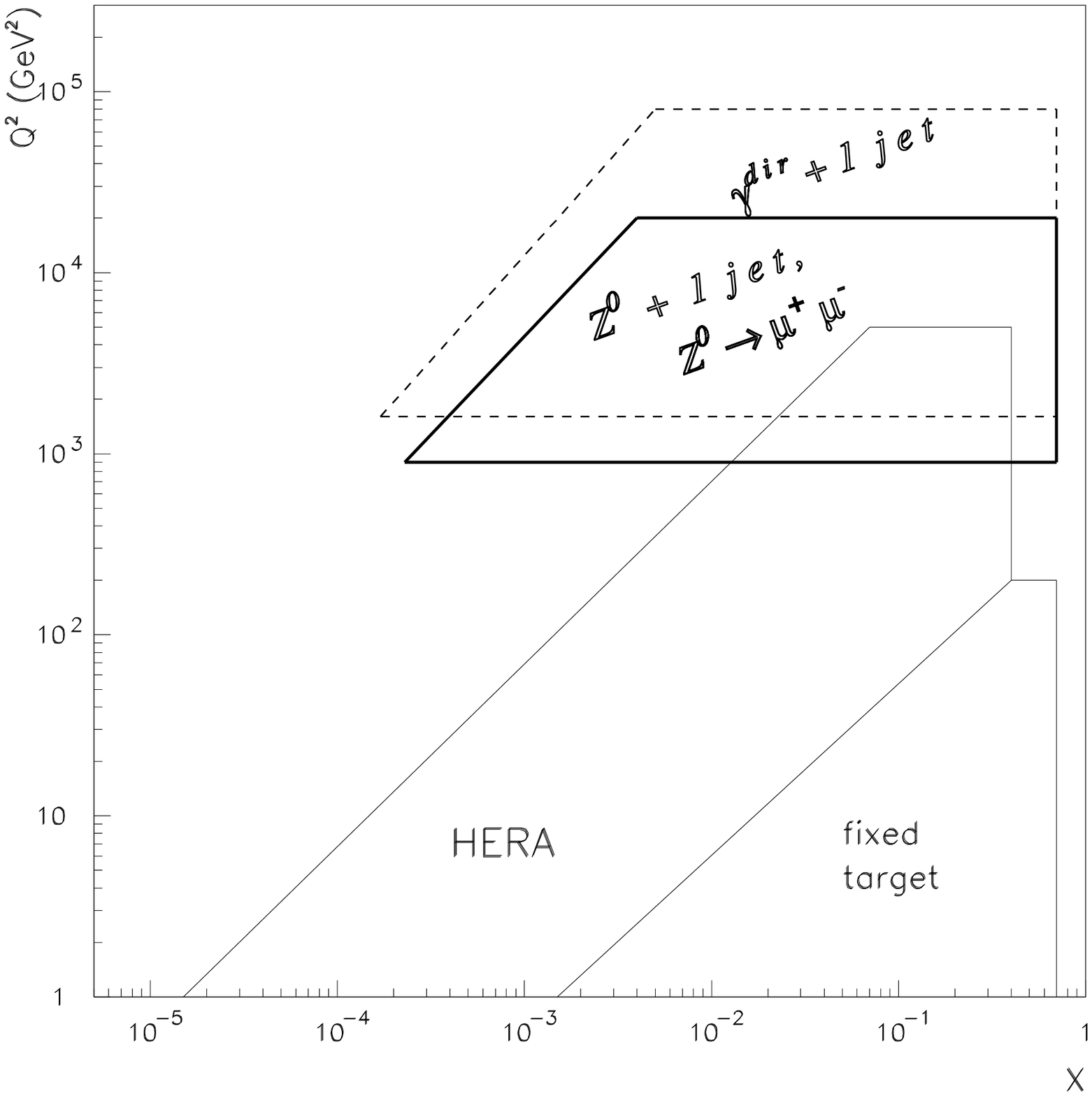}
\vspace*{-5.0mm}
\caption{LHC  $(x,Q^2)$ kinematic region for the process
$pp\to Z^0 + jet+X$ ~~(with $Z^0\to \mu^+\mu^-)$.}
\label{fig:zpj_xQ2}
  \vskip2mm
\end{figure}

 The area that can be covered by studying the process (\ref{eq:zpj}) with 
the subsequent decay $Z^0\to\mu^+\mu^-$ is shown in Fig.~\ref{fig:zpj_xQ2}.
The number of events in different $x$ and $Q^2$ intervals of
this area is given in Table \ref{tab:B30}.
From this figure (and Tables~\ref{tab:B30}) 
it is seen that during first two years of LHC running 
at low luminosity ($L=10^{33}\,cm^{-2} s^{-1}$) it would be possible to extract an information
for the gluon distribution determination $f^g_p(x,Q^2)$ in a proton in the region of 
$0.9\cdot10^3\leq Q^2\leq 4\cdot 10^4 ~(GeV/c)^2$ with as small $x$ values as accessible at HERA
but at higher $Q^2$ values (by 1--2 orders of magnitude).
It is also worth emphasizing that the  sample of the \zpj events selected for this aim can be used
to perform a cross-check of $f^g_p(x,Q^2)$ determination with help of \gpj events 
\cite{BKS_GLU,GLU_BKGD}.
The area covered with \gpj events is also shown in Fig.~\ref{fig:zpj_xQ2} by dashed lines.
\begin{table}[htbp]
\small
\begin{center}
\caption{Numbers of \zpj events (with $Z^0\to\mu^+\mu^-)$ in 
$Q^2$ and $x$ intervals for $L_{int}=20 ~fb^{-1}$.}
\label{tab:B30}
\vskip0.1cm
\begin{tabular}{|lc|r|r|r|r|r|c|}                  \hline
 & $Q^2$ &\multicolumn{4}{c|}{ \hspace{-0.9cm} $x$ values of a parton} &All $x$ 
&$\Pt^{Z}$   \\\cline{3-7}
 & $(GeV/c)^2$ & $10^{-4}$--$10^{-3}$ & $10^{-3}$--$10^{-2}$ &$10^{-2}$--
$10^{-1}$ & $10^{-1}$--$10^{0}$ & $10^{-4}$--$10^{0}$&$(GeV/c)$     \\\hline
&\hmm\hmm  900-1600\hmm  &18409  & 45844  & 47453  &  2479  & 114185& 30--40\\\hline 
&\hmm\hmm 1600-2500\hmm  & 7417  & 28361  & 28702  &  1854  & 66333 & 40--50\\\hline
&\hmm\hmm 2500-3600\hmm  & 2479  & 16574  & 19015  &  1533  & 39599 & 50--60\\\hline
&\hmm\hmm 3600-5000\hmm  & 1097  & 10406  & 12941  &  1533  & 25977 & 60--71 \\\hline
&\hmm\hmm 5000-6400\hmm  &  227  &  5846  &  6944  &  1022  & 14039 & 71--80\\\hline
&\hmm\hmm 6400-8100\hmm  &  170  &  4238  &  5430  &   624  & 10463 & 80--90\\\hline
&\hmm\hmm 8100-10000\hmm &   19  &  2989  &  4049  &   719  &  7776 & 90--100\\\hline
&\hmm\hmm 10000-14400\hmm&   19  &  2819  &  4579  &   908  &  8325 & 100--120\\\hline
&\hmm\hmm 14400-20000\hmm&    0  &  1400  &  2781  &   454  &  4635 & 120--141\\\hline
&\hmm\hmm 20000-40000\hmm&    0  &   908  &  2195  &   719  &  3822 & 141--200 \\\hline
\multicolumn{6}{c|}{}&{\bf 295~144}\\\cline{7-7}
\end{tabular}
\end{center}
\end{table}

\normalsize


%
\section{Summary.}
           
A possibility of the absolute jet energy scale setting with help of
\zpj events based on the $qg\to q+Z^0$ and $q\bar{q}\to g+Z^0$ subprocesses
with subsequent $Z^0$ decay to the muon pair is studied. 
The PYTHIA event generator is applied here to find the selection criteria of
the \zpj events that would provide a good $\Pt^{Z}-\Pt^{Jet}$ balance.

It is shown here (by analogy with \cite{BKS_P1}--\cite{BKS_P5}) that the limitation of the clusters $\Pt$
that may be found in an event in addition to the main jet as well as the limitation
 of $\Pt$ activity of all particles beyond the \zpj system (see Section 2)
leads to an improvement of the $\Ptz-\Pt^{Jet}$ balance value. 
A further improvement of the $\Ptz-\Pt^{Jet}$ balance can be reached by
selection of events having the isolated jets only.
Besides, this criterion (as well as the simultaneous jet finding by
two or three algorithms; see Selection 3 in Section 2.2) 
is also important for selecting events with a clean jet topology and 
rising the confidence level of a jet determination.
The summarizing results of our study of
the jet energy scale setting are presented in Appendices 2--6 (see also Fig.~\ref{fig:mu-sig}). 

It is demonstrated (Section 7) that the used selection criteria guarantee practically complete suppression of
the combinatorial background from the QCD events.

It is worth emphasizing  that the number of events    
presented here were not out main goal as they may depend on the used event
generator and on the particular choice of
a long set of its parameters. 
The most important result of our work is demonstration that the set of new selection criteria 
(limitation of $\Pt^{clust}$, $\Pt^{out}$ and jet isolation found earlier in \cite{BKS_P1}--\cite{BKS_P5})
are also very useful for the  jet energy scale determination by help of \zpj events.

It is also shown that the selected sample of the \zpj events,
most suitable for the absolute jet energy  scale setting at the LHC energy, 
can provide useful information
for the gluon density determination inside a proton in the kinematic region with
$x$ values as small as accessible at HERA  
but at much higher $Q^2$ values (by about 1--2 orders of magnitude): 
$2\cdot 10^{-4}\leq x \leq 1.0$ with $0.9\cdot10^3\leq Q^2\leq 4\cdot 10^4 ~(GeV/c)^2$.
This sample of \zpj events  can be used
to perform a cross-check of the $f^g_p(x,Q^2)$ determination by help of \gpj events 
\cite{BKS_GLU,GLU_BKGD}.
The $x-Q^2$ kinematic area that can be covered by the \zpj events (with $Z^0\to\mu^+\mu^-$)
as well as by the \gpj events is shown in Fig.~\ref{fig:zpj_xQ2}.


~\\[3mm]
{\Large\bf Acknowledgments.}
\normalsize
\rm
 
We are greatly thankful to D.~Denegri for having offered this theme to study,
fruitful discussions and permanent support and encouragement.
It is a pleasure for us
to express our recognition for helpful discussions to P.~Aurenche,
M.~Dittmar, M.~Fontannaz, J.Ph.~Guillet, M.L.~Mangano, E.~Pilon,
H.~Rohringer, S.~Tapprogge and especially to J.~Womersley for supplying us with
the preliminary version of paper [1].

\end{document}